\documentclass[11pt,sort&compress,twoside]{elsarticle}

\usepackage{amsmath,amssymb,graphicx,multirow,relsize,rotating,subfigure,tabularx}

\usepackage[margin=1.9cm]{geometry}

\usepackage{hyperref}
\hypersetup{%
bookmarksnumbered=true,
bookmarksopen=true,
hypertexnames=false,
pdfauthor={Fabio Clementi},
pdfkeywords={k-generalized statistics, wealth distribution, finite mixture models},
pdfpagemode=UseOutlines,
pdfstartview=FitH,
pdfsubject={arXiv.org e-print},
pdftitle={A generalized statistical model for the size distribution of wealth}
}

\usepackage[flushleft]{threeparttable}

\usepackage[nottoc]{tocbibind}

\journal{JSTAT}

\defcitealias{ClementiGallegatiKaniadakis2009}{J. Stat. Mech. (2009) P02037}

\begin{document}


\begin{frontmatter}

\title{\textbf{A generalized statistical model for the size distribution of wealth}}

\author[unimc]{F. Clementi\corref{cor}}
\ead{fabio.clementi@unimc.it}

\author[univpm]{M. Gallegati}
\ead{mauro.gallegati@univpm.it}

\author[polito]{G. Kaniadakis}
\ead{giorgio.kaniadakis@polito.it}

\cortext[cor]{Corresponding author. Tel.: +39--0733--258--3962; fax: +39--0733--258--3970}

\address[unimc]{Dipartimento di Scienze Politiche, della Comunicazione e delle Relazioni Internazionali, Universit\`a degli Studi di Macerata, Piazza G. Oberdan 3, 62100 Macerata, Italy}

\address[univpm]{Dipartimento di Scienze Economiche e Sociali, Universit\`a Politecnica delle Marche, Piazzale R. Martelli 8, 60121 Ancona, Italy}

\address[polito]{Dipartimento Scienza Applicata e Tecnologia, Politecnico di Torino, Corso Duca degli Abruzzi 24, 10129 Torino, Italy}

\begin{abstract}
In a recent paper in this journal [\citetalias{ClementiGallegatiKaniadakis2009}] we proposed a new, physically motivated, distribution function for modeling individual incomes having its roots in the framework of the $\kappa$-generalized statistical mechanics. The performance of the $\kappa$-generalized distribution was checked against real data on personal income for the United States in 2003. In this paper we extend our previous model so as to be able to account for the distribution of wealth. Probabilistic functions and inequality measures of this generalized model for wealth distribution are obtained in closed form. In order to check the validity of the proposed model, we analyze the U.S. household wealth distributions from 1984 to 2009 and conclude an excellent agreement with the data that is superior to any other model already known in the literature.
\end{abstract}

\end{frontmatter}


\section{Introduction}

The quantitative and formal development of the personal or size distribution of income and the measurement of income inequality was first introduced by the Italian economist Vilfredo Pareto. He specified his type I model early in 1895 \cite{Pareto1895}, and in 1896 and 1897 his types II and III \cite{Pareto1896,Pareto1897a,Pareto1897b}, and made an inequality interpretation of his shape parameter. Based on Pareto's economic foundations, and on the stochastic foundations afterward developed by other authors \cite{Mandelbrot1960,Ord1975}, the Pareto law (Pareto type I) is now overwhelmingly considered as the income distribution model of high income groups.

After Pareto's seminal contribution, many probability density functions have been proposed in the literature that are suitable for describing the size distribution of income amongst the population as a whole\textemdash see e.g. the comprehensive survey contained in \cite{KleiberKotz2003}. Fitting of parametric functional forms has also been common for the distribution of \textit{wealth}.\footnote{Income and wealth are commonly used to assess the economic well-being of individuals, families or households. Although some correlation exists between them, the relationship is not perfect: greater income is likely to mean greater wealth, but not always. The two measures, in fact, are not synonymous. Income is a \textit{flow}, since it is meaningful only when defined in relation to a period of time (hourly, weekly, monthly or annual income). Wealth is a \textit{stock}, increasing as new assets are acquired or savings accumulated, and the only time information required is when the stock was measured (no periodicity is necessary). The link between the flow from income and the stock of wealth is obvious: the greater the former, the more rapidly the latter will increase. Accordingly, a high income may be associated with low wealth\textemdash this is the case, for example, with young people starting their careers; on the other hand, a low income could accompany high wealth\textemdash this is the case with some retirees who have little income but who have accumulated and paid for substantial assets. At a practical level, wealth is distributed much more unequally than income because of life cycle savings and bequest motives \cite{DaviesSandstromShorrocksWolff2009}. Data on stocks of wealth also present distinctive features in comparison with income data that make empirical analysis non-standard in several ways (see the ongoing discussion above for details). However, as far as the shape of the particular distribution is concerned, income and wealth share qualitatively the same characteristics: many empirical wealth distributions are indeed positively skewed with ``fat'' and long right-hand tails, as are income distributions.} However, the problem for the wealth researcher is that virtually all of the models suggested within the context of the income distribution literature are defined for variables taking only strictly positive values, although published statistical data of wealth distributions give clear evidence of presenting highly significant frequencies of households or individuals with null and/or negative wealth. The early contributions systematically dismissed these frequencies and fitted their respective proposed models to the positive observations only, thus omitting a significant part of the story.\footnote{In the 1950s, Refs. \cite{WoldWhittle1957} and \cite{Sargan1957} proposed the Pareto type I model and the lognormal distribution, respectively. Afterward, other models were proposed: in 1969 the Pareto types I and II by \cite{Stiglitz1969}; in 1975, the log-logistic by \cite{Atkinson1975} and the Pearson type V by \cite{Vaughan1975}. All of these models are restricted to describe only the positive range of wealth, since they are not defined for zero and/or negative values.}

To the best of our knowledge, Dagum \cite{Dagum1977,Dagum1978} was the first and only one to specify and test a four-parameter model for wealth distributions (Dagum type II). The fourth parameter in the Dagum model is an estimate of the frequency of economic units with wealth equal to zero. This model is highly relevant to describe total (gross) wealth distribution because of the always large observed percentage of economic units with null total wealth. Dagum \cite{Dagum1990,Dagum1994,Dagum2006a,Dagum2006b} made further developments of his type II model to analyze the distribution of \textit{net} wealth, which is equal to gross wealth minus total debt. The support of the Dagum model of net wealth is the real line $\mathbb{R}=\left(-\infty,\infty\right)$, thus allowing to fit the subset of economic units with null and negative wealth. Furthermore, it contains as particular cases both the Dagum types I and II distributions \cite{Dagum1977}.

More in detail, the Dagum general model of net wealth distribution is a mixture (or a convex combination) of an atomic and two continuous distributions. The atomic distribution concentrates its unit mass of economic agents at zero, and therefore accounts for the economic units with null net wealth. The continuous distribution accounting for the negative net wealth observations is given by a Weibull function. It has a fast left tail convergence to zero, and therefore it has finite moments of all orders. The other continuous distribution, specified as the Dagum type I model, accounts for the positive values of net wealth and presents a heavy right tail, thus having a small number of finite moments of positive order. This different behavior at the two tails of the distribution stems form the fact that, unlike the right tail of income and (gross or net) wealth distributions\textemdash which tend slowly to zero when income and wealth tend to infinity, the distribution of the negative values (left tail) of net wealth tends very fast to zero when the variable tends to minus infinity, since economic units face a short term challenge of either moving out of the negative range of net wealth or bankruptcy.

The purpose of the present work is to provide estimates for the 1984--2009 U.S. net wealth distributions of this Dagum general model, partly motivated by the fact that there are no applications other than Dagum's ones \cite{Dagum1990,Dagum1994,Dagum2006a,Dagum2006b} that we are aware of\textemdash the only notable exception being represented by \cite{JenkinsJantti2005}, who fitted the model to Finnish net wealth data in 1984 and 1989. Furthermore, since other approaches can be entertained and comparative study of their relative merits performed, we also explore the possibility of using alternative distributions to characterize positive net wealth values. That is, we formalize, analyze and fit to our U.S. net wealth data finite mixture models based upon the Singh-Maddala and $\kappa$-generalized distributions as specifications for the positive values. The Singh-Maddala \cite{SinghMaddala1976} is known to be very successful in fitting the empirical income distributions. The $\kappa$-generalized was proposed in previous works of us \cite{ClementiGallegatiKaniadakis2007,ClementiDiMatteoGallegatiKaniadakis2008,ClementiGallegatiKaniadakis2009,ClementiGallegatiKaniadakis2010,ClementiGallegatiKaniadakis2012} to describe the distribution of personal income in some developed economies for different years. Positive conclusions were drawn about its ability to provide an accurate description of the observed distributions, ranging from the low to the middle region, and up to the right tail. The empirical success of the $\kappa$-generalized was complemented by goodness-of-fit comparisons showing that fitting the distribution to available income data offers superior performance over other existing models (including the Singh-Maddala and Dagum type I) in a significant number of cases.

The content of the paper is organized as follows: Section \ref{sec:TheKappaGeneralizedStatisticalDistributionAndItsProperties} recalls some basic properties of the $\kappa$-generalized statistical distribution; Section \ref{sec:SpecificationOfFiniteMixtureModelsForNetWealthDistribution} presents the main analytical properties of the net wealth distribution models; Section \ref{sec:MomentsOfFiniteMixtureModelsForNetWealth} deduces their corresponding moments; Section \ref{sec:TheLorenzCurveAndTheGiniRatioOfTheNetWealthDistributionModels} derives the parametric forms of the Lorenz curve and Gini ratio for the distribution of net wealth; Section \ref{sec:Application} fits the specified models to the U.S. data on household net wealth covering the years 1984--2009; and Section \ref{sec:SummaryAndConclusions} presents the conclusions.


\section{\texorpdfstring{The $\kappa$-generalized statistical distribution and its properties}{The k-generalized statistical distribution and its properties}}
\label{sec:TheKappaGeneralizedStatisticalDistributionAndItsProperties}

After 2001, a physical mechanism emerging in the context of special relativity was proposed by one of us \cite{Kaniadakis2001a,Kaniadakis2001b,Kaniadakis2002,Kaniadakis2005}, predicting a deformation of the exponential function. According to this mechanism, the classical exponential distribution transforms into a new distribution, which at high energies presents a Pareto fat tail. More precisely, this mechanism deforms the ordinary exponential function $\exp\left(x\right)$ into the generalized exponential function $\exp_{\kappa}\left(x\right)$ given by
\begin{equation}
\exp_{\kappa}\left(x\right)=\left(\sqrt{1+\kappa^{2}x^{2}}+\kappa x\right)^{\frac{1}{\kappa}}.
\end{equation}
The above deformation is generated by the fact that the propagation of the information has a finite speed, and the deformation parameter $\kappa$ is proportional to the reciprocal of this speed. The $\kappa$-generalized exponential has the important properties
\begin{subequations}
\begin{equation}
\exp_{\kappa}\left(x\right){\atop\stackrel{\textstyle\sim}{\scriptstyle x\rightarrow\pm\infty}}\left|2\kappa x\right|^{\pm\frac{1}{\left|\kappa\right|}},
\label{eq:ParetoLaw}
\end{equation}
\begin{equation}
\exp_{\kappa}\left(x\right){\atop\stackrel{\textstyle\sim}{\scriptstyle x\rightarrow\,0}}\exp\left(x\right).
\label{eq:ExponentialDistribution}
\end{equation}
\end{subequations}

It is remarkable that for classical systems where the information propagates instantaneously it results $\kappa=0$, so that the ordinary exponential emerges naturally after noting that $\exp_{0}\left(x\right)=\exp\left(x\right)$. Moreover, in the low energy region $x\rightarrow0$ according to Eq. \eqref{eq:ExponentialDistribution} the exponential distribution emerges again, because the system behaves classically. On the contrary, in systems where the information propagates with a finite speed\textemdash these systems are intrinsically relativistic\textemdash it results $\kappa\neq0$, so that the exponential tails become fat according to Eq. \eqref{eq:ParetoLaw} and the Pareto law emerges.

The generalized exponential represents a very useful and powerful tool to formulate a new statistical theory capable to treat systems described by distribution functions exhibiting power-law tails and admitting a stable entropy \cite{Kaniadakis2009a,Kaniadakis2009b}. Furthermore, non-linear evolution models already known in statistical physics \cite{KaniadakisDelsanto1993,KaniadakisLavagnoQuarati1996,KaniadakisLavagnoQuarati1997} can be easily adapted or generalized within the new theory.

The function $\exp_{\kappa}\left(x\right)$ was also adopted successfully in the analysis of various non physical systems \cite{RajaonarisonBolducJayet2005,Rajaonarison2008,BertottiModanese2012}. In Refs. \cite{ClementiGallegatiKaniadakis2007,ClementiDiMatteoGallegatiKaniadakis2008,ClementiGallegatiKaniadakis2009,ClementiGallegatiKaniadakis2010,ClementiGallegatiKaniadakis2012} we have used the function $\exp_{\kappa}\left(x\right)$ to model the personal income distribution by defining the cumulative distribution function through
\begin{equation}
F\left(x\right)=1-\exp_{\kappa}\left[-\left(x/\beta\right)^{\alpha}\right],\quad x\geq0,\quad\alpha,\beta>0,\quad\kappa\in[0,1).
\end{equation}
The corresponding probability density function reads
\begin{equation}
f\left(x\right)=
\frac{\alpha}{\beta}\left(\frac{x}{\beta}\right)^{\alpha-1}\frac{\exp_{\kappa}\left[-\left(\frac{x}{\beta}\right)^{\alpha}\right]}{\sqrt{1+\kappa^{2}\left(\frac{x}{\beta}\right)^{2\alpha}}}.
\label{eq:KappaDistribution}
\end{equation}
It follows immediately that for low incomes the distribution function behaves similarly to the Weibull model $F\left(x\right)=1-\exp\left[-\left(x/\beta\right)^{\alpha}\right]$, whereas for large $x$ it approaches a Pareto distribution with scale $\beta\left(2\kappa\right)^{-\frac{1}{\alpha}}$ and shape $\frac{\alpha}{\kappa}$, i.e. $F\left(x\right){\atop\stackrel{\textstyle\sim}{\scriptstyle x\rightarrow+\infty}}1-\left[\frac{\beta\left(2\kappa\right)^{-\frac{1}{\alpha}}}{x}\right]^{\frac{\alpha}{\kappa}}$. Similarly, the density function for $x\rightarrow0^{+}$ behaves as a Weibull distribution $f\left(x\right)=\frac{\alpha}{\beta}\left(\frac{x}{\beta}\right)^{\alpha-1}\exp\left[-\left(x/\beta\right)^{\alpha}\right]$, while for $x\rightarrow+\infty$ it reduces to the Pareto's law
$f\left(x\right)=\frac{\frac{\alpha}{\kappa}\left[\beta\left(2\kappa\right)^{-\frac{1}{\alpha}}\right]^{\frac{\alpha}{\kappa}}}{x^{\frac{\alpha}{\kappa}+1}}$.


\section{Specification of finite mixture models for net wealth distribution}
\label{sec:SpecificationOfFiniteMixtureModelsForNetWealthDistribution}

The general model of net wealth distribution as a mixture of an atomic and two continuous distributions takes the form
\begin{equation}
f\left(w\right)=\sum^{3}_{i=1}\theta_{i}f_{i}\left(w\right),\quad -\infty<w<\infty,\quad\theta_{i}\geq0,\quad\sum_{i}\theta_{i}=1,
\label{eq:Equation1}
\end{equation}
where $w$ denotes the wealth variable and $\left\{\theta_{i}\right\}_{i=1,\ldots,3}$ are the mixture proportions. The two-parameter Weibull density
\begin{equation}
f_{1}\left(w\right)=\frac{s}{\lambda}\left(\frac{\left|w\right|}{\lambda}\right)^{s-1}\exp\left[-\left(\frac{\left|w\right|}{\lambda}\right)^{s}\right],\quad w<0,\quad \left(s,\lambda\right)>0
\label{eq:Equation2}
\end{equation}
describes the distribution of economic units with negative net wealth, while the null net wealth observations are accounted for by a distribution that concentrates its unit mass at $w=0$, i.e.
\begin{equation}
f_{2}\left(0\right)=1.
\label{eq:Equation3}
\end{equation}
The other continuous distribution, $f_{3}\left(w\right)$, accounts for the positive values of net wealth, and is alternatively specified by the following three-parameter densities:
\begin{enumerate}
\item the Singh-Maddala
\begin{equation}
f^{\mathrm{SM}}_{3}\left(w\right)=\frac{aqw^{a-1}}{b^{a}\left[1+\left(\frac{w}{b}\right)^{a}\right]^{1+q}},\quad w>0,\quad\left(a,b,q\right)>0;
\label{eq:Equation4}
\end{equation}
\item the Dagum type I
\begin{equation}
f^{\mathrm{D}}_{3}\left(w\right)=\frac{apw^{ap-1}}{b^{ap}\left[1+\left(\frac{w}{b}\right)^{a}\right]^{p+1}},\quad w>0,\quad\left(a,b,p\right)>0;
\label{eq:Equation5}
\end{equation}
\item the $\kappa$-generalized given by Eq. \eqref{eq:KappaDistribution}.
\end{enumerate}

The corresponding cumulative distribution function reads
\begin{equation}
F\left(w\right)=\theta_{1}F_{1}\left(w\right)+\theta_{2}F_{2}\left(w\right)+\theta_{3}F_{3}\left(w\right),\quad \theta_{1}+\theta_{2}=\rho,\quad \theta_{3}=1-\rho,
\label{eq:Equation7}
\end{equation}
where
\begin{subequations}
\begin{equation}
F_{1}\left(w\right)=
\begin{cases}
\exp\left[-\left(\frac{\left|w\right|}{\lambda}\right)^{s}\right],\quad w<0;\\
1,\quad w\geq 0;
\end{cases}
\label{eq:Equation8a}
\end{equation}
\begin{equation}
F_{2}\left(w\right)=
\begin{cases}
0,\quad w<0;\\
1,\quad w\geq 0;
\end{cases}
\label{eq:Equation8b}
\end{equation}
\begin{equation}
F_{3}\left(w\right)=
\begin{cases}
0,\quad w\leq 0;\\
F_{3}\left(w\right),\quad w>0.
\end{cases}
\label{eq:Equation8c}
\end{equation}
\label{eq:Equation8}
\end{subequations}
Hence
\begin{equation}
F\left(w\right)=
\begin{cases}
\theta_{1}\exp\left[-\left(\frac{\left|w\right|}{\lambda}\right)^{s}\right],\quad w<0;\\
\rho,\quad w=0;\\
\rho+\left(1-\rho\right)F_{3}\left(w\right),\quad w>0,
\end{cases}
\label{eq:Equation9}
\end{equation}
with $F_{3}\left(w\right)$ having the following alternative mathematical specifications
\begin{subequations}
\begin{equation}
F^{\mathrm{SM}}_{3}\left(w\right)=1-\left[1+\left(\frac{w}{b}\right)^{a}\right]^{-q};\\
\label{eq:Equation10a}
\end{equation}
\begin{equation}
F^{\mathrm{D}}_{3}\left(w\right)=\left[1+\left(\frac{w}{b}\right)^{-a}\right]^{-p};\\
\label{eq:Equation10b}
\end{equation}
\begin{equation}
F^{\kappa\text{-gen}}_{3}\left(w\right)=1-\exp_{\kappa}\left[-\left(\frac{w}{\beta}\right)^{\alpha}\right].
\label{eq:Equation10c}
\end{equation}
\label{eq:Equation10}
\end{subequations}


\section{Moments of finite mixture models for net wealth distribution}
\label{sec:MomentsOfFiniteMixtureModelsForNetWealth}

It follows from model \eqref{eq:Equation1} that the $r$th-order moment about the origin is\footnote{In what follows, $\Gamma\left(\cdot\right)$ stands for the Euler gamma function.}
\begin{equation}
\mu_{r}=E\left(W^{r}\right)=\int\limits^{\infty}_{-\infty}w^{r}f\left(w\right)\operatorname{d}w=\theta_{1}E_{1}\left(W^{r}\right)+\theta_{2}E_{2}\left(W^{r}\right)+\theta_{3}E_{3}\left(W^{r}\right),
\label{eq:Equation11}
\end{equation}
where
\begin{equation}
E_{1}\left(W^{r}\right)=(-1)^{r}\lambda^{r}\Gamma\left(1+\frac{r}{s}\right)
\label{eq:Equation12}
\end{equation}
and $E_{2}\left(W^{r}\right)=0$.

As for $E_{3}\left(W^{r}\right)$ in the last member of Eq. \eqref{eq:Equation11}, according to the alternative distributions considered to characterize positive net wealth values one gets\footnote{See \cite{KleiberKotz2003} for relevant expressions. Formulas for the moments of the $\kappa$-generalized distribution are given in \cite{ClementiDiMatteoGallegatiKaniadakis2008,ClementiGallegatiKaniadakis2009,ClementiGallegatiKaniadakis2010,ClementiGallegatiKaniadakis2012}.}
\begin{subequations}
\begin{equation}
E^{\mathrm{SM}}_{3}\left(W^{r}\right)=\frac{b^{r}\Gamma\left(1+\frac{r}{a}\right)\Gamma\left(q-\frac{r}{a}\right)}{\Gamma\left(q\right)};
\label{eq:Equation13a}
\end{equation}
\begin{equation}
E^{\mathrm{D}}_{3}\left(W^{r}\right)=\frac{b^{r}\Gamma\left(p+\frac{r}{a}\right)\Gamma\left(1-\frac{r}{a}\right)}{\Gamma\left(p\right)};
\label{eq:Equation13b}
\end{equation}
\begin{equation}
E^{\kappa\text{-gen}}_{3}\left(W^{r}\right)=\beta^{r}\left(2\kappa\right)^{-\frac{r}{\alpha}}\frac{\Gamma\left(1+\frac{r}{\alpha}\right)}{1+\frac{r}{\alpha}\kappa}\frac{\Gamma\left(\frac{1}{2\kappa}-\frac{r}{2\alpha}\right)}{\Gamma\left(\frac{1}{2\kappa}+\frac{r}{2\alpha}\right)}.
\label{eq:Equation13c}
\end{equation}
\label{eq:Equation13}
\end{subequations}

The mean net wealth equals
\begin{equation}
\mu_{1}=E\left(W\right)=-\theta_{1}\lambda\Gamma\left(1+\frac{1}{s}\right)+\theta_{3}E_{3}\left(W\right),
\label{eq:Equation14}
\end{equation}
where $E_{3}\left(W\right)$ is alternatively given by Eqs. \eqref{eq:Equation13} with $r=1$.


\section{The Lorenz curve and the Gini ratio of the net wealth distribution models}
\label{sec:TheLorenzCurveAndTheGiniRatioOfTheNetWealthDistributionModels}

By definition, the Lorenz curve \cite{Lorenz1905} describes a relation between the cumulative distribution function, $F\left(w\right)$, and the first cumulative moment distribution function, given by
\begin{equation}
L\left(u\right)=\frac{1}{\mu_{1}}\int\limits^{w}_{0}w^{'}f\left(w^{'}\right)\operatorname{d}w^{'}=\frac{1}{\mu_{1}}\int\limits^{u}_{0}w\left(u^{'}\right)\operatorname{d}u^{'},\quad u\in\left[0,1\right],
\label{eq:Equation15}
\end{equation}
where $u=F\left(w\right)$ and $w\left(u\right)=F^{-1}\left(u\right)$ denotes the quantile function. Given the mathematical structure of the general net wealth distribution model \eqref{eq:Equation1} and \eqref{eq:Equation7}, we have
\begin{subequations}
\begin{align}
L^{\mathrm{SM}}\left(u\right)&=
\begin{cases}
\mathlarger{-\frac{\lambda\theta_{1}}{\mu_{1}}\Gamma\left(1+\frac{1}{s},\log\frac{\theta_{1}}{u}\right)},\quad 0\leq u<\theta_{1};\\[2ex]
\mathlarger{-\frac{\lambda\theta_{1}}{\mu_{1}}\Gamma\left(1+\frac{1}{s}\right)},\quad \theta_{1}\leq u\leq\rho;\\[2ex]
\begin{aligned}
&\frac{1}{\mu_{1}}\left\{\left(1-\rho\right)bq\left[B\left(q-\frac{1}{a},1+\frac{1}{a}\right)-B\left(\left[\frac{1-u}{1-\rho}\right]^{\frac{1}{q}};q-\frac{1}{a},1+\frac{1}{a}\right)\right]\right.\\[1ex]
&\left.-\lambda\theta_{1}\Gamma\left(1+\frac{1}{s}\right)\right\},\quad u>\rho;
\end{aligned}
\end{cases}
\label{eq:Equation19a}\\[1ex]
L^{\mathrm{D}}\left(u\right)&=
\begin{cases}
\mathlarger{-\frac{\lambda\theta_{1}}{\mu_{1}}\Gamma\left(1+\frac{1}{s},\log\frac{\theta_{1}}{u}\right)},\quad 0\leq u<\theta_{1};\\[2ex]
\mathlarger{-\frac{\lambda\theta_{1}}{\mu_{1}}\Gamma\left(1+\frac{1}{s}\right)},\quad \theta_{1}\leq u\leq\rho;\\[2ex]
\mathlarger{\frac{1}{\mu_{1}}\Biggl\{\left(1-\rho\right)bpB\left(\left[\frac{u-\rho}{1-\rho}\right]^{\frac{1}{p}};p+\frac{1}{a},1-\frac{1}{a}\right)-\lambda\theta_{1}\Gamma\left(1+\frac{1}{s}\right)\Biggr\}},\quad u>\rho;
\end{cases}
\label{eq:Equation19b}\\[1ex]
L^{\kappa\text{-gen}}\left(u\right)&=
\begin{cases}
\mathlarger{-\frac{\lambda\theta_{1}}{\mu_{1}}\Gamma\left(1+\frac{1}{s},\log\frac{\theta_{1}}{u}\right)},\quad 0\leq u<\theta_{1};\\[2ex]
\mathlarger{-\frac{\lambda\theta_{1}}{\mu_{1}}\Gamma\left(1+\frac{1}{s}\right)},\quad \theta_{1}\leq u\leq\rho;\\[2ex]
\begin{aligned}
&\frac{1}{\mu_{1}}\left\{\frac{\left(1-\rho\right)\beta}{\left(2\kappa\right)^{1+\frac{1}{\alpha}}}\left[B\left(\frac{1}{2\kappa}-\frac{1}{2\alpha},1+\frac{1}{\alpha}\right)-B\left(\left[\frac{1-u}{1-\rho}\right]^{2\kappa};\frac{1}{2\kappa}-\frac{1}{2\alpha},1+\frac{1}{\alpha}\right)\right]\right.\\[1ex]
&\left.-\lambda\theta_{1}\Gamma\left(1+\frac{1}{s}\right)\right\},\quad u>\rho,
\end{aligned}
\end{cases}
\label{eq:Equation19c}
\end{align}
\label{eq:Equation19}
\end{subequations}
\hspace{-3pt}where $B\left(\cdot,\cdot\right)$ and $B\left(\cdot;\cdot,\cdot\right)$ denote, respectively, the complete and incomplete Euler beta functions. Eqs. \eqref{eq:Equation19} determine the path of the net wealth Lorenz curve $L\left(u\right)$ over the closed interval $\left[0,1\right]$ for the different specifications of the net wealth finite mixture model. It follows that for $u=1$, $L\left(1\right)=1$.

Since the net wealth Lorenz curve presents negative values for all $u<\rho$, it can be proved that the Gini inequality ratio takes the form \cite{Dagum2006a,Dagum2006b}
\begin{equation}
G=\left\{2\int\limits^{1}_{0}\left[u-L\left(u\right)\right]\operatorname{d}u\right\}/\left[1+\rho\left|L\left(\theta_{1}\right)\right|\right]=\left[1-2\int\limits^{1}_{0}L\left(u\right)\operatorname{d}u\right]/\left[1-\rho L\left(\theta_{1}\right)\right],
\label{eq:Equation20}
\end{equation}
where
\begin{equation}
\int\limits^{1}_{0}L\left(u\right)\operatorname{d}u=\int\limits^{\theta_{1}}_{0}L\left(u\right)\operatorname{d}u+\int\limits^{\rho}_{\theta_{1}}L\left(u\right)\operatorname{d}u+\int\limits^{1}_{\rho}L\left(u\right)\operatorname{d}u.
\label{eq:Equation21}
\end{equation}

Using Eqs. \eqref{eq:Equation19}, the Gini ratio becomes
\begin{subequations}
\begin{equation}
G^{\mathrm{SM}}=\frac{\mu_{1}-2\left[\left(1-\rho\right)^{2}bqB\left(2q-\frac{1}{a},1+\frac{1}{a}\right)-\lambda\theta_{1}\left(1-\theta_{1}2^{-1-\frac{1}{s}}\right)\Gamma\left(1+\frac{1}{s}\right)\right]}{\mu_{1}+\rho\lambda\theta_{1}\Gamma\left(1+\frac{1}{s}\right)};
\label{eq:Equation25a}
\end{equation}
\begin{equation}
G^{\mathrm{D}}=\frac{\mu_{1}-2\left\{\left(1-\rho\right)^{2}bp\left[B\left(p+\frac{1}{a},1-\frac{1}{a}\right)-B\left(2p+\frac{1}{a},1-\frac{1}{a}\right)\right]-\lambda\theta_{1}\left(1-\theta_{1}2^{-1-\frac{1}{s}}\right)\Gamma\left(1+\frac{1}{s}\right)\right\}}{\mu_{1}+\rho\lambda\theta_{1}\Gamma\left(1+\frac{1}{s}\right)};
\label{eq:Equation25b}
\end{equation}
\begin{equation}
G^{\kappa\text{-gen}}=\frac{\mu_{1}-2\left[\frac{\left(1-\rho\right)^{2}\beta}{\left(2\kappa\right)^{1+\frac{1}{\alpha}}}B\left(\frac{1}{\kappa}-\frac{1}{2\alpha},1+\frac{1}{\alpha}\right)-\lambda\theta_{1}\left(1-\theta_{1}2^{-1-\frac{1}{s}}\right)\Gamma\left(1+\frac{1}{s}\right)\right]}{\mu_{1}+\rho\lambda\theta_{1}\Gamma\left(1+\frac{1}{s}\right)}.
\label{eq:Equation25c}
\end{equation}
\label{eq:Equation25}
\end{subequations}


\section{Application}
\label{sec:Application}


\subsection{The U.S. data on household net wealth}

The empirical analysis is based on data drawn from the Panel Study of Income Dynamics (PSID), a nationally representative household survey collected by the Survey Research Center at the University of Michigan since 1968. The PSID provides detailed information about economic, demographic, sociological and psychological aspects of many U.S. households. Since the focus is on the distribution of wealth, we use all (nine) waves currently available of the special PSID supplement asking information on household wealth holdings. This supplement was added in 1984 and was conducted on a periodic basis prior to 1999 (in 1984, 1989 and 1994). After 1997 the basic PSID survey switched to biennial data collection, and starting with 1999 the wealth questions have been included in each wave (1999, 2001, 2003, 2005, 2007 and 2009).

As shown in Table \ref{tab:Table1}, the number of households participating in the various waves varies between 6 and 9 thousand, providing samples for analysis that are reasonably representative of the ``true'' wealth distribution in the U.S.\footnote{For more on this issue, see for instance \cite{Hill1991} and \cite{Wolff1999}. In particular, measured against the standards set by two prominent American household wealth surveys\textemdash the Survey of Consumer Finances (SCF) and the Survey of Income and Program Participation (SIPP)\textemdash the PSID does not differ substantially from them when it comes to measuring total wealth and its distribution among the great bulk of the U.S. population. Moreover, its measurement error characteristics look to be consistently better than are those of the SCF and the SIPP: the PSID has indeed a lower item nonresponse rate than these alternative data sets and thus less need to construct imputed values \cite{CurtinJusterMorgan1989}.} In particular, we are concerned with the distribution of \textit{net} wealth, which is constructed as sum of values of several asset types net of debt held by each household.\footnote{The PSID asks about eight broad wealth categories: (1) value of farm or business assets; (2) value of checking and savings accounts, money market funds, certificates of deposit, savings bonds, Treasury bills, other Individual Retirement Accounts (IRAs); (3) value of real estate other than main home; (4) value of shares of stock in publicly held corporations, mutual funds or investment trusts, including stocks in IRAs; (5) value of vehicles or other assets ``on wheels''; (6) value of other investments in trusts or estates, bond funds, life insurance policies, special collections; (7) value of private annuities or IRAs; (8) value of home equity (calculated as home value minus remaining mortgage). More complete definitions of the asset and debt categories are available at the PSID web site: \url{http://psidonline.isr.umich.edu/}.} Since net wealth is expressed in nominal local currency units, all figures have been deflated to allow for meaningful comparisons over the period covered by the data. To do so, we have employed the Consumer Price Index deflator (yearly series based on year 2005) provided by the OECD.\footnote{Available at: \url{http://stats.oecd.org/}.} Furthermore, after a simple adjustment for differences in relative needs of households according to their size,\footnote{When the distribution of wealth is defined over households and not over individuals, a problem arises with regard to the possibility of comparing wealth holdings of different units. The reason is that households vary in size and thus wealth levels are not a good indicator of their well-being, as households with a different number of members may have different needs in the use of wealth even when this is the same order of magnitude across them. In this case, a correction should be made to meaningfully compare different situations. This correction is called an \textit{equivalence scale}. There is a wide range of equivalence scales in use in different countries and by different organizations. All take account of household size: in many scales this is the only factor, whilst in those taking into account other considerations it is the factor with greatest weight. Choices of equivalence scale in recent wealth studies are reviewed in \cite{SierminskaSmeeding2005}. Here we adopt a simple equivalence scale that is most commonly used in international studies \cite{AtkinsonRainwaterSmeeding1995} where net household wealth is divided by the square root of the number of household members.} net wealth values have been weighted by using appropriate sampling weights provided by the PSID staff in order to produce representative estimates for all households in the target population.

Table \ref{tab:Table1} also provides a number of summary statistics.
%
\begin{table}[!t]
\centering
\begin{threeparttable}
\caption{Summary statistics for U.S. household net wealth, 1984--2009}
\newcolumntype{R}{>{\raggedleft\arraybackslash}X}%
\small
\begin{tabularx}{\textwidth}{lRRRRRRRRr}
\hline\hline
\multirow{2}{*}{Stats}&\multicolumn{9}{c}{Wave}\\
\cline{2-10}
&1984&1989&1994&1999&2001&2003&2005&2007&2009\\
\hline
Obs&6,918&7,113&7,415&6,851&7,195&7,565&8,002&8,289&8,690\\
Mean&121,613&135,095&135,885&185,055&189,139&201,991&223,506&256,281&248,753\\
Median&36,940&38,988&42,390&45,958&50,735&50,295&53,276&56,744&39,143\\
Skewness&18.340&15.592&13.821&18.234&19.673&15.079&17.883&26.006&31.766\\
Kurtosis&410.821&364.775&302.102&454.349&598.511&317.070&513.130&909.552&1,193.073\\
Gini&0.758&0.759&0.751&0.789&0.774&0.788&0.782&0.803&0.850\\
$\text{\% with } W<0$&6.807&8.096&8.636&9.363&9.409&9.439&10.240&11.137&14.385\\
$\text{\% with } W=0$&4.285&4.554&4.751&3.557&3.428&3.872&3.715&3.725&4.484\\
$\text{\% with } W>0$&88.908&87.350&86.614&87.080&87.163&86.689&86.045&85.138&81.130\\
\hline\hline
\end{tabularx}
\begin{tablenotes}
\footnotesize
\item\textit{Source}: Authors' own calculations using the PSID supplemental wealth files.
\end{tablenotes}
\label{tab:Table1}
\end{threeparttable}
\end{table}
%
Consider first the prevalence of zero and negative values. On the basis of the PSID data, the proportion of households with negative net wealth rose steadily between 1984 and 2009 (from less than 7\% to over 14\%) whilst the proportion of households with zero net wealth increased somewhat between 1984 and 1994 (from slightly more than 4\% to about 5\%) followed by a decline towards 3\% until 2001. By 2003, the percentage of these households started increasing again to almost 4\% and stayed nearly the same in the following two waves (2005 and 2007) before reaching, in 2009, about the same level of 1984. Notwithstanding these differences in the proportions of negatives and zeros with regard to time trends and levels, when their joint prevalence is taken into account we find it to be relatively high on average (around 14\% of the sample size). This situation is quite different from that generally faced in the case of income data, where it is often assumed that income can only take on positive values\textemdash in practice, there may be non-positive incomes but usually the number of these is so small that one can just ignore them. By contrast, in the case of net wealth data the assumption of dealing with a positive quantity can not be justified, since it is a matter of fact that many people enter a period of indebtedness at some point in their life. Therefore, net wealth may legitimately take on negative and zero values, and the proportion of such observations could be non-negligible (as in our case) in representative samples of the target population.\footnote{For futher discussion on this issue, we refer the reader to \cite{JenkinsJantti2005} and \cite{Cowell2011}.}

Results on time trend in real mean household net wealth show that it rose continuously by some 111\% from 1984 to 2007 and then fell by almost 3\% between 2007 and 2009, for an overall annual growth rate of about 3\% over the entire period. The time trend (although not the magnitude of level changes) in median net wealth appears to mirror that of the mean. Indeed, the PSID data show median net wealth rising in real terms by some 54\% from 1984 to 2007\textemdash save for a temporary slight decrease by less than 1\% between 2001 and 2003\textemdash and then quickly reaching the same level as in 1989 by a sharp fall-off of around 31\% between 2007 and 2009, for an overall annual growth rate of about 0.2\% over the twenty-five years.

\begin{figure}[!t]
\centering
\includegraphics[width=1.00\textwidth]{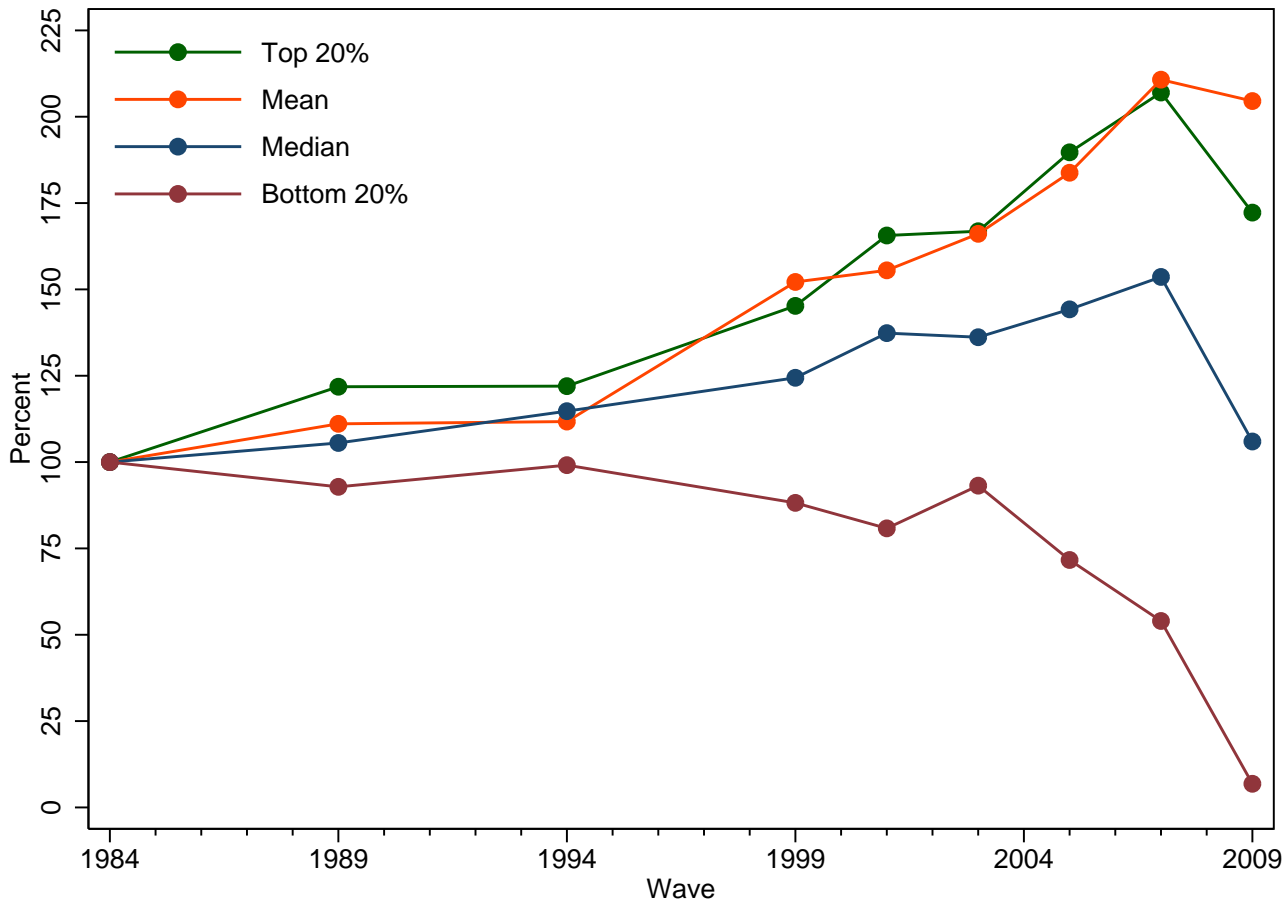}
\caption{Statistics of U.S. household net wealth distribution, 1984--2009}
\label{fig:Figure1}
\end{figure}

The change over time in the relationship between the mean and median is shown in Figure \ref{fig:Figure1}. To provide an indication of how the distribution of wealth across households has changed, the evolution of the relative positions of households at the two ends of the distribution (i.e, the bottom and top quintile groups or bottom and top 20\%) is also displayed.\footnote{Changes in the aggregates of Figure \ref{fig:Figure1} over the twenty-five year span are measured by \textit{index numbers}. An index number is calculated by dividing the value in the year of interest by the value in the base year\textemdash 1984 in our case\textemdash and then multiplying the result by 100. The base year index is always 100 and the index for each subsequent year will be above or below 100, depending on whether there as been an increase or decrease in the data compared with the base year.}
%
%
As noted above, both mean and median net wealth increased from 1984 to 2007, with the mean typically increasing to a greater extent than the median. This suggests that in recent decades wealth became more concentrated among households at the upper end of the distribution, and indeed in those years where the divergence between the mean and the median became wider\textemdash i.e., between 1994 and 2007\textemdash the largest changes in net wealth holdings of households in the top of the real distribution were also observed. By contrast, both measures fell during the 2007--2009 recession. The relatively greater decline in the median than in the mean suggests that the recession more adversely affected the households in the bottom of the wealth distribution than those further up, as shown by the worsening relative position for the bottom 20\% of them.

One might suspect that the differences in the pace of real growth between the mean and median net wealth are partly caused by the presence of long and heavy tails in the distribution of U.S. household net wealth, particularly at the top of the data range. Indeed, the positive skewness values listed in the fourth row of Table \ref{tab:Table1} suggest that the distribution of net wealth in any one year has a long tail toward the upper end, thus indicating a non-trivial prevalence of values that are ``extremes'' in relation to the rest of the data. Furthermore, in each of the wave years the level of kurtosis is huge as compared to the normal distribution (fifth row of Table \ref{tab:Table1}), meaning that the upper tail of net wealth distribution is inevitably ``fat''\textemdash i.e. declines to zero more slowly than exponentially. As the median would not be affected by the extreme values, this results in average net wealth holdings that are consistently larger than median ones in all cases.

\begin{figure}[!t]
\centering
\includegraphics[width=1.00\textwidth]{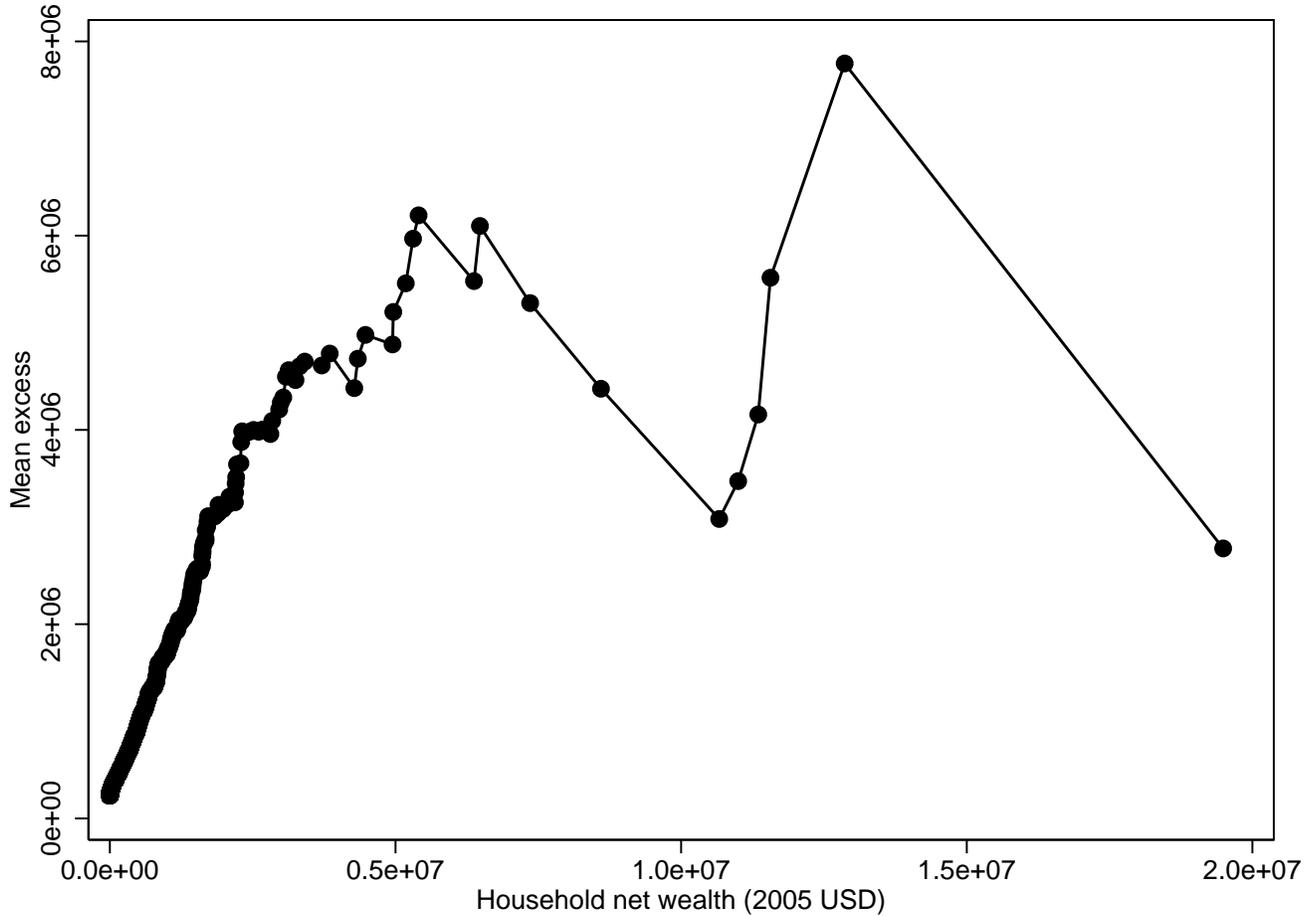}
\caption{Mean excess plot for the positive values of U.S. household net wealth in 2003}
\label{fig:Figure2}
\end{figure}

Additional information about the fatness of the upper tail of the U.S. net wealth distribution can be obtained from visual examination of the sample mean excess plot shown in Figure \ref{fig:Figure2}.\footnote{Properties of the mean excess plot are reviewed, for instance, in \cite{BeirlantGoegebeurTeugelsSegers2004}. We do not report plots for each year but they are available upon request. Since we are interested here in the upper tail behavior of the distribution, the plot has been drawn only for the positive values of net wealth.}
%
%
For a sequence of threshold values $\left\{w_{i}\right\}_{i=1,\ldots,N}$, the mean excess plot reports the mean of exceedances over $w_{i}$ against $w_{i}$ itself. Putting it differently, this is a plot of the set of pairs $\left(w_{i},e_{n}\left(w_{i}\right)\right)_{i=1,\ldots,N-1}$, where $e_{n}\left(w_{i}\right)=\frac{1}{\sum^{N}_{j=i+1}\pi_{j}}\sum^{N}_{j=i+1}\pi_{j}\left(w_{j}-w_{i}\right)$ is the sample \textit{mean excess function} (weighted by household weights $\left\{\pi_{i}\right\}_{i=1,\ldots,N}$) and $\left\{w_{i}\right\}_{i=1,\ldots,N}$ are the sample observations ranked from least to greatest. If the points in the plot show an upward trend, then this is a sign of heavy-tailed behavior. Exponentially distributed data would give an approximately horizontal line and data from a short-tailed distribution would show a downward trend. In particular, if the empirical mean excess plot seems to follow a reasonably straight line with positive slope above a certain net wealth value, then this is an indication of Pareto (power-law) behavior in tail. This is precisely the kind of behavior we observe in the 2003 PSID data. In fact, apart from some noisiness by the most extreme observations, there is evidence for consistent upward trends of the data and straightening out of the plots above some points onwards, hence providing a statistical justification for the emergence of power laws as limiting behavior for the very wealthy.

\begin{figure}[!t]
\centering
\includegraphics[width=1.00\textwidth]{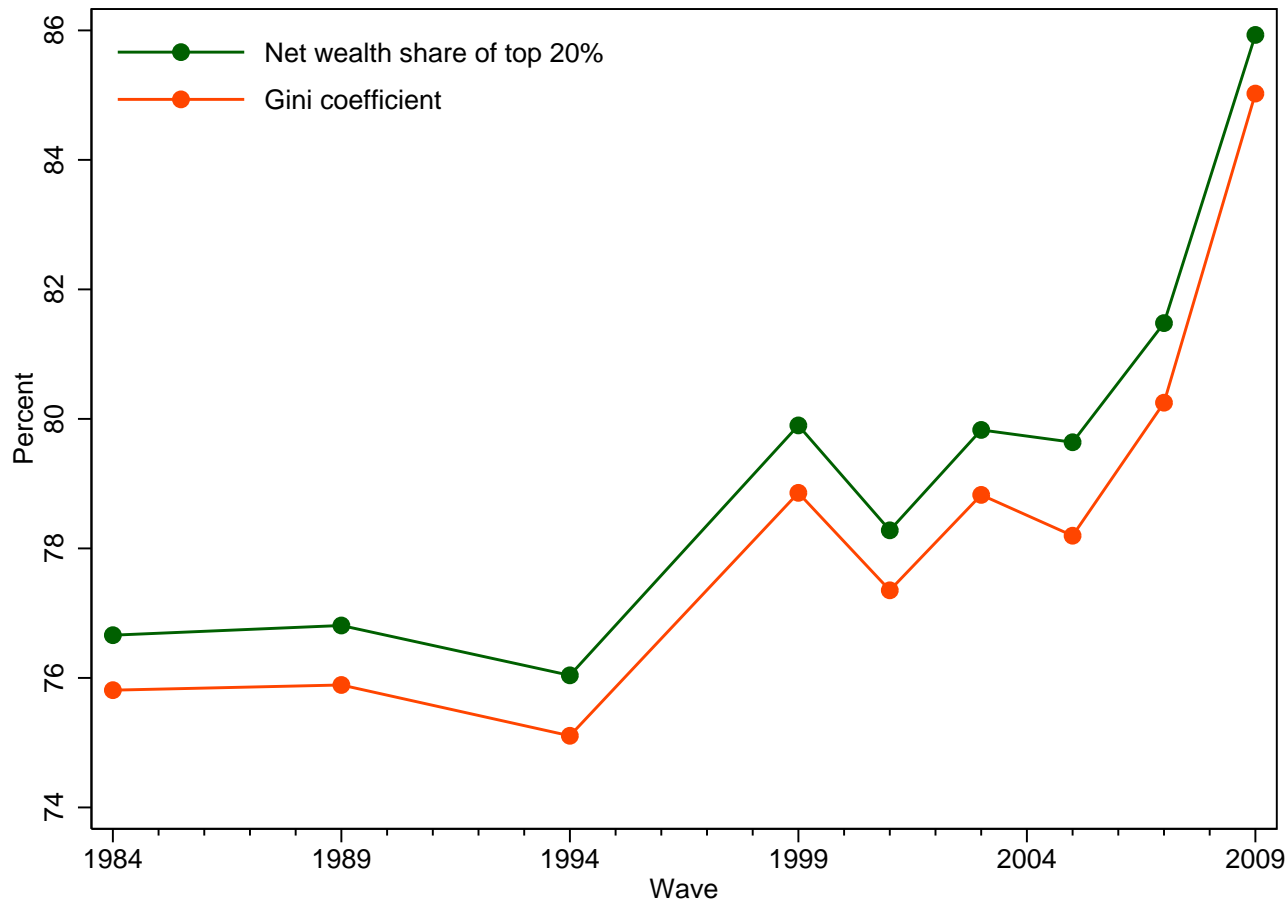}
\caption{Gini coefficient and net wealth share of top 20\% across years}
\label{fig:Figure3}
\end{figure}

Does this finding matter when it comes to inequality judgments? Figure \ref{fig:Figure3} displays the pattern of Gini coefficient for the distribution of U.S. household net wealth over the period 1984--2009 (the corresponding values are reported in the fourth-last row of Table \ref{tab:Table1}).
%
%
At least three different sub-periods are shown: from the second half of the 1980s to the first half of the 1990s, from the late 1990s to the first half of the 2000s and the last time interval (2007--2009). According to the PSID, net wealth inequality remained virtually unchanged during the first sub-period. Indeed, the Gini coefficient rose slightly between 1984 and 1989 (from 0.758 to 0.759) and then fell in 1994 to a level below that of 1984 (0.751). By contrast, inequality increased sharply between 1994 and 1999, with the Gini coefficient of net wealth climbing to 0.789. The following years still show almost the same degree of inequality: the Gini coefficient was estimated at 0.788 in 2003 and 0.782 in 2005, except for a temporary decrease to a value of 0.774 in 2001. Finally, between 2007 and 2009 net wealth inequality was up steeply, with the Gini coefficient advancing from 0.803 to 0.850.

Figure \ref{fig:Figure3} also displays the evolution between 1984 and 2009 of the share of total net wealth held by the richest 20\% of households, which amounted on average to around 80\% of the whole over the period. A noteworthy result is that the observed time pattern of inequality seems to have been driven by the conspicuous wealth holdings at the very top end of the distribution. Indeed, as can be seen from the figure, the time profile of net wealth share of the wealthiest 20\% is analogous to that of Gini coefficient: after rising to a peak in 1999, it went down and then started to increase again until 2009.\footnote{The correlation coefficient between the two series of Gini coefficient and the net wealth share received by the top 20\% is 0.998, which is highly significant ($p\text{-value}<0.001$).}

To sum up the above, wealth in the U.S. has become more concentrated in recent decades. Net wealth inequality increased by the mid-1990s, and the increase was not interrupted during the 2007--2009 recession. The share of total net wealth held by the top wealth owners has also grown during the same period, whereas at the other end of the wealth distribution there was a sharp increase in the number of households with zero or negative net wealth. Needless to say, this has resulted in a widening gap between the rich and the poor that advocates more attention be paid to the implementation of appropriate and practical policies aimed at reducing inequalities, limiting their negative effects on the socio-economic system and reversing the mechanisms producing them \cite{Stiglitz2012}.


\subsection{Estimation and comparison of finite mixture models for net wealth distribution}

Table \ref{tab:Table2} presents the parameter estimates and other relevant statistics arising from the fitting of the net wealth distribution models previously discussed to the PSID data from 1984 to 2009.
%
\begin{sidewaystable}[p]
\centering
\begin{threeparttable}
\caption{Estimated mixture models for the U.S. household net wealth, 1984--2009\tnote{a}}
\small
\begin{tabular}{llrrrrrrrrrrrrr}
\hline\hline
\multirow{2}{*}{Wave}&\multirow{2}{*}{Model}&\multicolumn{8}{c}{Parameters\tnote{b}}&\multirow{2}{*}{logLik}&\multirow{2}{*}{AIC}&\multirow{2}{*}{BIC}&\multirow{2}{*}{Mean\tnote{d}}&\multirow{2}{*}{Gini\tnote{e}}\\
\cline{3-10}
&&$a$ ($\alpha$)&$b$ ($\beta$)&$q,p,\kappa$&$\gamma$\tnote{c}&$\theta_{1}$&$\theta_{2}$&$s$&$\lambda$&&\\
\hline
\multirow{6}{*}{1984}&\multirow{2}{*}{SM}&0.757&373,565&3.754&\multirow{2}{*}{2.843}&0.068&0.043&0.578&4,511&\multirow{2}{*}{-84,249}&\multirow{2}{*}{168,511}&\multirow{2}{*}{168,579}&\multirow{2}{*}{111,616}&\multirow{2}{*}{0.736}\\
&&(0.012)&(56,465)&(0.316)&&(0.003)&(0.002)&(0.018)&(382)&&&&\\
&\multirow{2}{*}{D}&1.614&138,163&0.377&\multirow{2}{*}{0.608}&0.068&0.043&0.578&4,511&\multirow{2}{*}{-84,230}&\multirow{2}{*}{168,474}&\multirow{2}{*}{168,542}&\multirow{2}{*}{121,361}&\multirow{2}{*}{0.753}\\
&&(0.042)&(5,720)&(0.015)&&(0.003)&(0.002)&(0.018)&(382)&&&&\\
&\multirow{2}{*}{$\kappa$-gen}&0.718&76,514&0.374&\multirow{2}{*}{1.919}&0.068&0.043&0.578&4,511&\multirow{2}{*}{-84,229}&\multirow{2}{*}{168,471}&\multirow{2}{*}{168,539}&\multirow{2}{*}{114,181}&\multirow{2}{*}{0.741}\\
&&(0.009)&(1,663)&(0.022)&&(0.003)&(0.002)&(0.018)&(382)&&&&\\
\hline
\multirow{6}{*}{1989}&\multirow{2}{*}{SM}&0.743&459,094&3.814&\multirow{2}{*}{2.833}&0.081&0.046&0.619&6,639&\multirow{2}{*}{-87,573}&\multirow{2}{*}{175,161}&\multirow{2}{*}{175,230}&\multirow{2}{*}{130,298}&\multirow{2}{*}{0.749}\\
&&(0.012)&(76,568)&(0.348)&&(0.003)&(0.002)&(0.018)&(473)&&&&\\
&\multirow{2}{*}{D}&1.520&152,781&0.402&\multirow{2}{*}{0.611}&0.081&0.046&0.619&6,639&\multirow{2}{*}{-87,583}&\multirow{2}{*}{175,180}&\multirow{2}{*}{175,249}&\multirow{2}{*}{152,050}&\multirow{2}{*}{0.781}\\
&&(0.040)&(7,050)&(0.017)&&(0.003)&(0.002)&(0.018)&(473)&&&&\\
&\multirow{2}{*}{$\kappa$-gen}&0.702&89,241&0.367&\multirow{2}{*}{1.912}&0.081&0.046&0.619&6,639&\multirow{2}{*}{-87,565}&\multirow{2}{*}{175,143}&\multirow{2}{*}{175,212}&\multirow{2}{*}{133,595}&\multirow{2}{*}{0.754}\\
&&(0.009)&(1,998)&(0.023)&&(0.003)&(0.002)&(0.018)&(473)&&&&\\
\hline
\multirow{6}{*}{1994}&\multirow{2}{*}{SM}&0.769&449,157&3.713&\multirow{2}{*}{2.857}&0.086&0.048&0.716&10,759&\multirow{2}{*}{-91,866}&\multirow{2}{*}{183,745}&\multirow{2}{*}{183,816}&\multirow{2}{*}{133,116}&\multirow{2}{*}{0.745}\\
&&(0.012)&(72,386)&(0.338)&&(0.003)&(0.002)&(0.020)&(629)&&&&&\\
&\multirow{2}{*}{D}&1.528&154,125&0.421&\multirow{2}{*}{0.644}&0.086&0.048&0.716&10,759&\multirow{2}{*}{-91,879}&\multirow{2}{*}{183,771}&\multirow{2}{*}{183,842}&\multirow{2}{*}{156,209}&\multirow{2}{*}{0.779}\\
&&(0.038)&(6,831)&(0.017)&&(0.003)&(0.002)&(0.020)&(629)&&&&&\\
&\multirow{2}{*}{$\kappa$-gen}&0.727&95,745&0.379&\multirow{2}{*}{1.919}&0.086&0.048&0.716&10,759&\multirow{2}{*}{-91,861}&\multirow{2}{*}{183,736}&\multirow{2}{*}{183,807}&\multirow{2}{*}{137,029}&\multirow{2}{*}{0.751}\\
&&(0.010)&(2,083)&(0.024)&&(0.003)&(0.002)&(0.020)&(629)&&&&&\\
\hline
\multirow{6}{*}{1999}&\multirow{2}{*}{SM}&0.724&477,324&3.380&\multirow{2}{*}{2.446}&0.094&0.036&0.751&11,529&\multirow{2}{*}{-86,534}&\multirow{2}{*}{173,082}&\multirow{2}{*}{173,152}&\multirow{2}{*}{173,472}&\multirow{2}{*}{0.775}\\
&&(0.012)&(77,394)&(0.288)&&(0.004)&(0.002)&(0.022)&(642)&&&&&\\
&\multirow{2}{*}{D}&1.422&181,486&0.420&\multirow{2}{*}{0.597}&0.094&0.036&0.751&11,529&\multirow{2}{*}{-86,548}&\multirow{2}{*}{173,111}&\multirow{2}{*}{173,181}&\multirow{2}{*}{212,146}&\multirow{2}{*}{0.812}\\
&&(0.038)&(9,213)&(0.018)&&(0.004)&(0.002)&(0.022)&(642)&&&&&\\
&\multirow{2}{*}{$\kappa$-gen}&0.678&107,494&0.389&\multirow{2}{*}{1.742}&0.094&0.036&0.751&11,529&\multirow{2}{*}{-86,527}&\multirow{2}{*}{173,067}&\multirow{2}{*}{173,137}&\multirow{2}{*}{179,416}&\multirow{2}{*}{0.781}\\
&&(0.009)&(2,579)&(0.024)&&(0.004)&(0.002)&(0.022)&(642)&&&&&\\
\hline
\multirow{6}{*}{2001}&\multirow{2}{*}{SM}&0.683&980,104&4.669&\multirow{2}{*}{3.188}&0.094&0.034&0.724&11,083&\multirow{2}{*}{-91,364}&\multirow{2}{*}{182,742}&\multirow{2}{*}{182,812}&\multirow{2}{*}{181,487}&\multirow{2}{*}{0.765}\\
&&(0.010)&(197,710)&(0.486)&&(0.003)&(0.002)&(0.020)&(623)&&&&&\\
&\multirow{2}{*}{D}&1.514&229,564&0.366&\multirow{2}{*}{0.554}&0.094&0.034&0.724&11,083&\multirow{2}{*}{-91,373}&\multirow{2}{*}{182,759}&\multirow{2}{*}{182,829}&\multirow{2}{*}{211,855}&\multirow{2}{*}{0.794}\\
&&(0.041)&(10,520)&(0.015)&&(0.003)&(0.002)&(0.020)&(623)&&&&&\\
&\multirow{2}{*}{$\kappa$-gen}&0.652&118,900&0.326&\multirow{2}{*}{2.004}&0.094&0.034&0.724&11,083&\multirow{2}{*}{-91,354}&\multirow{2}{*}{182,722}&\multirow{2}{*}{182,792}&\multirow{2}{*}{185,349}&\multirow{2}{*}{0.769}\\
&&(0.008)&(2,778)&(0.023)&&(0.003)&(0.002)&(0.020)&(623)&&&&&\\
\hline\hline
\end{tabular}
\label{tab:Table2}
\end{threeparttable}
\end{sidewaystable}
%
\setcounter{table}{1}
\begin{sidewaystable}[p]
\centering
\begin{threeparttable}
\caption{continued\tnote{a}}
\small
\begin{tabular}{llrrrrrrrrrrrrr}
\hline\hline
\multirow{2}{*}{Wave}&\multirow{2}{*}{Model}&\multicolumn{8}{c}{Parameters\tnote{b}}&\multirow{2}{*}{logLik}&\multirow{2}{*}{AIC}&\multirow{2}{*}{BIC}&\multirow{2}{*}{Mean\tnote{d}}&\multirow{2}{*}{Gini\tnote{e}}\\
\cline{3-10}
&&$a$ ($\alpha$)&$b$ ($\beta$)&$q,p,\kappa$&$\gamma$\tnote{c}&$\theta_{1}$&$\theta_{2}$&$s$&$\lambda$&&\\
\hline
\multirow{6}{*}{2003}&\multirow{2}{*}{SM}&0.703&667,612&3.767&\multirow{2}{*}{2.649}&0.094&0.039&0.682&13,602&\multirow{2}{*}{-96,151}&\multirow{2}{*}{192,315}&\multirow{2}{*}{192,385}&\multirow{2}{*}{192,121}&\multirow{2}{*}{0.777}\\
&&(0.011)&(112,537)&(0.329)&&(0.003)&(0.002)&(0.019)&(791)&&&&&\\
&\multirow{2}{*}{D}&1.443&214,742&0.400&\multirow{2}{*}{0.577}&0.094&0.039&0.682&13,602&\multirow{2}{*}{-96,158}&\multirow{2}{*}{192,329}&\multirow{2}{*}{192,399}&\multirow{2}{*}{232,063}&\multirow{2}{*}{0.812}\\
&&(0.037)&(10,101)&(0.016)&&(0.003)&(0.002)&(0.019)&(791)&&&&&\\
&\multirow{2}{*}{$\kappa$-gen}&0.665&120,624&0.370&\multirow{2}{*}{1.796}&0.094&0.039&0.682&13,602&\multirow{2}{*}{-96,140}&\multirow{2}{*}{192,294}&\multirow{2}{*}{192,364}&\multirow{2}{*}{198,801}&\multirow{2}{*}{0.784}\\
&&(0.008)&(2,775)&(0.022)&&(0.003)&(0.002)&(0.019)&(791)&&&&&\\
\hline
\multirow{6}{*}{2005}&\multirow{2}{*}{SM}&0.660&1,371,330&4.978&\multirow{2}{*}{3.285}&0.102&0.037&0.728&11,804&\multirow{2}{*}{-102,470}&\multirow{2}{*}{204,954}&\multirow{2}{*}{205,024}&\multirow{2}{*}{216,342}&\multirow{2}{*}{0.775}\\
&&(0.010)&(290,880)&(0.531)&&(0.003)&(0.002)&(0.019)&(600)&&&&&\\
&\multirow{2}{*}{D}&1.456&267,209&0.371&\multirow{2}{*}{0.541}&0.102&0.037&0.728&11,804&\multirow{2}{*}{-102,482}&\multirow{2}{*}{204,978}&\multirow{2}{*}{205,048}&\multirow{2}{*}{264,727}&\multirow{2}{*}{0.813}\\
&&(0.037)&(12,103)&(0.015)&&(0.003)&(0.002)&(0.019)&(600)&&&&&\\
&\multirow{2}{*}{$\kappa$-gen}&0.632&138,871&0.315&\multirow{2}{*}{2.008}&0.102&0.037&0.728&11,804&\multirow{2}{*}{-102,462}&\multirow{2}{*}{204,937}&\multirow{2}{*}{205,008}&\multirow{2}{*}{221,517}&\multirow{2}{*}{0.780}\\
&&(0.008)&(3,206)&(0.022)&&(0.003)&(0.002)&(0.019)&(600)&&&&&\\
\hline
\multirow{6}{*}{2007}&\multirow{2}{*}{SM}&0.645&1,568,538&4.995&\multirow{2}{*}{3.221}&0.111&0.037&0.670&13,715&\multirow{2}{*}{-106,821}&\multirow{2}{*}{213,657}&\multirow{2}{*}{213,728}&\multirow{2}{*}{239,641}&\multirow{2}{*}{0.787}\\
&&(0.009)&(324,452)&(0.508)&&(0.003)&(0.002)&(0.015)&(713)&&&&&\\
&\multirow{2}{*}{D}&1.450&303,012&0.360&\multirow{2}{*}{0.523}&0.111&0.037&0.670&13,715&\multirow{2}{*}{-106,833}&\multirow{2}{*}{213,679}&\multirow{2}{*}{213,750}&\multirow{2}{*}{291,719}&\multirow{2}{*}{0.821}\\
&&(0.037)&(13,682)&(0.014)&&(0.003)&(0.002)&(0.015)&(713)&&&&&\\
&\multirow{2}{*}{$\kappa$-gen}&0.617&150,343&0.307&\multirow{2}{*}{2.007}&0.111&0.037&0.670&13,715&\multirow{2}{*}{-106,810}&\multirow{2}{*}{213,633}&\multirow{2}{*}{213,704}&\multirow{2}{*}{244,070}&\multirow{2}{*}{0.791}\\
&&(0.007)&(3,469)&(0.021)&&(0.003)&(0.002)&(0.015)&(713)&&&&&\\
\hline
\multirow{6}{*}{2009}&\multirow{2}{*}{SM}&0.640&930,909&3.988&\multirow{2}{*}{2.550}&0.144&0.045&0.707&17,847&\multirow{2}{*}{-110,594}&\multirow{2}{*}{221,202}&\multirow{2}{*}{221,273}&\multirow{2}{*}{221,273}&\multirow{2}{*}{0.828}\\
&&(0.009)&(171,313)&(0.348)&&(0.004)&(0.002)&(0.014)&(756)&&&&&\\
&\multirow{2}{*}{D}&1.334&245,091&0.393&\multirow{2}{*}{0.524}&0.144&0.045&0.707&17,847&\multirow{2}{*}{-110,605}&\multirow{2}{*}{221,225}&\multirow{2}{*}{221,295}&\multirow{2}{*}{295,004}&\multirow{2}{*}{0.869}\\
&&(0.033)&(12,039)&(0.015)&&(0.004)&(0.002)&(0.014)&(756)&&&&&\\
&\multirow{2}{*}{$\kappa$-gen}&0.605&128,792&0.353&\multirow{2}{*}{1.713}&0.144&0.045&0.707&17,847&\multirow{2}{*}{-110,583}&\multirow{2}{*}{221,181}&\multirow{2}{*}{221,251}&\multirow{2}{*}{230,967}&\multirow{2}{*}{0.835}\\
&&(0.007)&(3,120)&(0.022)&&(0.004)&(0.002)&(0.014)&(756)&&&&&\\
\hline\hline
\end{tabular}
\begin{tablenotes}
\footnotesize
\item\textit{Notes}: (a) SM = Singh-Maddala mixture model; D = Dagum mixture model; $\kappa$-gen = $\kappa$-generalized mixture model. (b) Numbers in round brackets: estimated standard errors. (c) Values of the Paretian upper tail index derived from parameter estimates of the Singh-Maddala (SM), Dagum (D) and $\kappa$-generalized ($\kappa$-gen) mixture models, respectively, as $\gamma^{\mathrm{SM}}=aq$, $\gamma^{\mathrm{D}}=ap$ and $\gamma^{\kappa\text{-gen}}=\frac{\alpha}{\kappa}$. (d) Analytic values obtained by substituting the estimated parameters into Eqs. \eqref{eq:Equation13} and \eqref{eq:Equation14} with $r=1$. (e) Analytic values obtained by substituting the estimated parameters into Eqs. \eqref{eq:Equation25}.
\item\textit{Source}: Authors' own calculations using the PSID supplemental wealth files.
\end{tablenotes}
\end{threeparttable}
\end{sidewaystable}
%
The parameters were estimated in all cases by minimizing the negative of the log-likelihood function via a modified Newton-Raphson procedure implemented in Stata's \texttt{ml} command \cite{StataCorp2011}, with the parameter covariance matrix estimates based on the negative inverse Hessian. Convergence was achieved easily within several iterations.

\begin{figure}[!t]
\centering
\subfigure[Mean]{\label{fig:Mean}\includegraphics[width=0.48\textwidth]{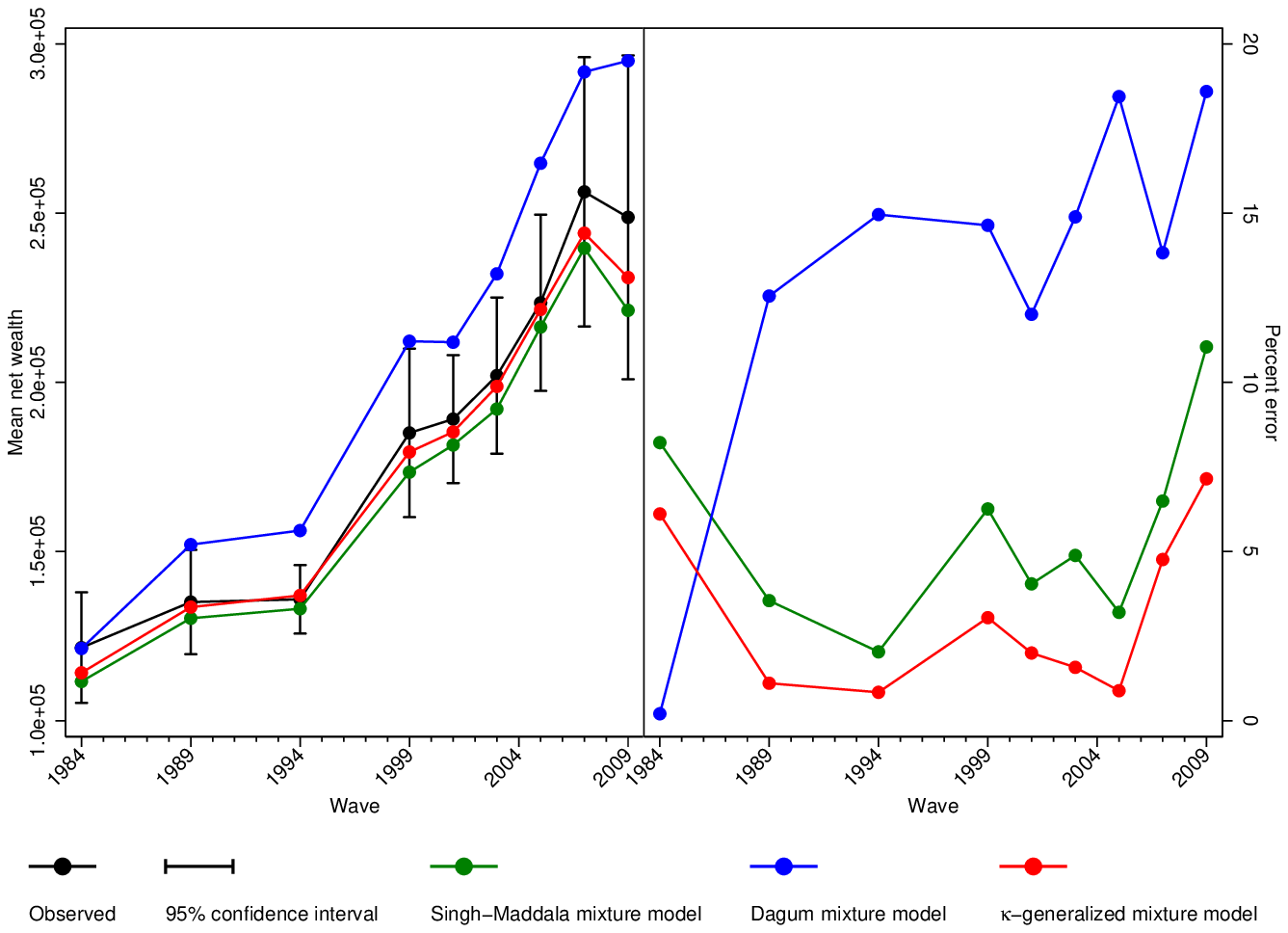}}
\quad
\subfigure[Gini]{\label{fig:Gini}\includegraphics[width=0.48\textwidth]{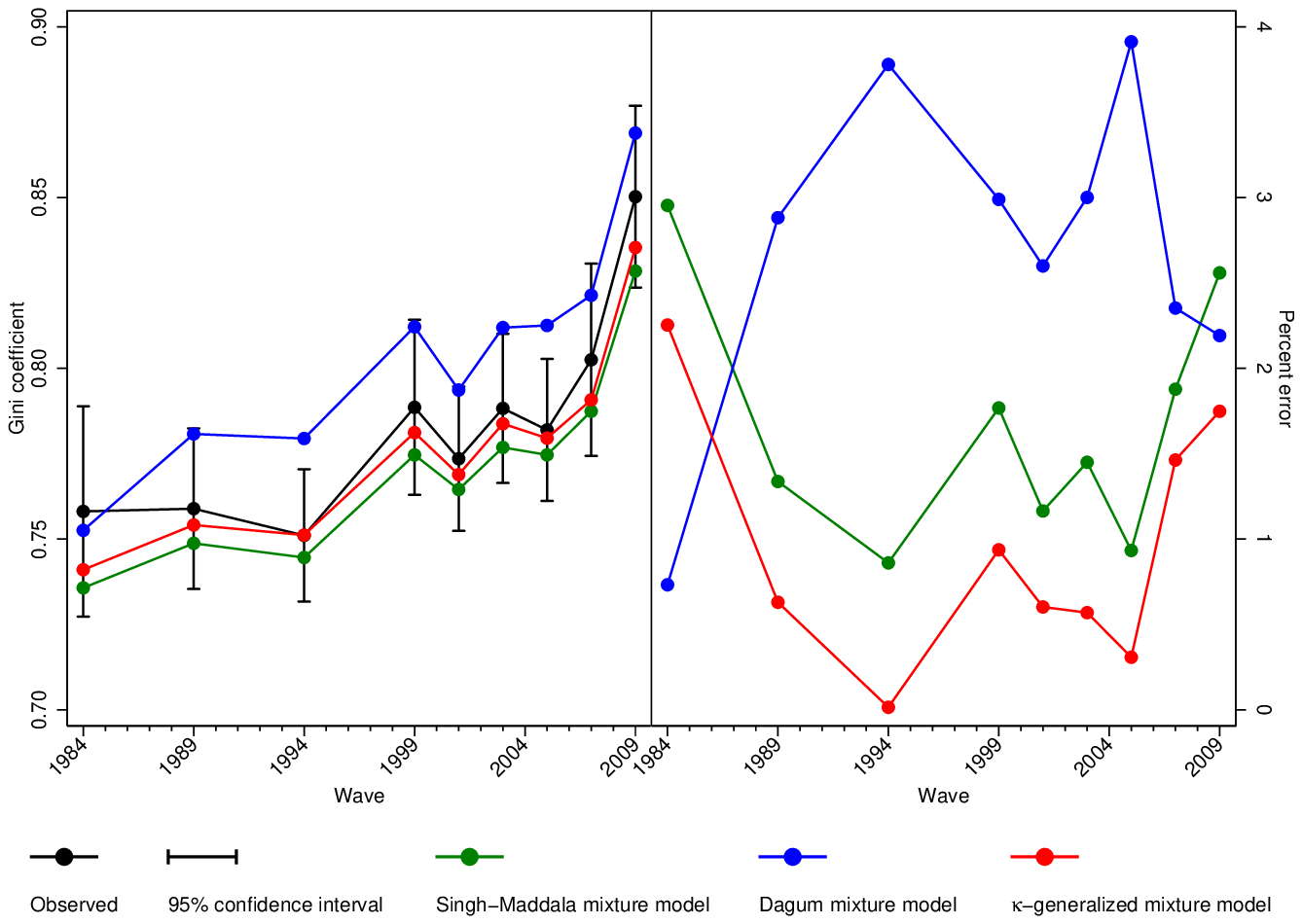}}
\caption{Observed and predicted values for the mean and Gini coefficient of U.S. household net wealth, 1984--2009. The vertical bars denote the symmetric 95\% normal-approximation confidence intervals for the empirical values calculated via the bootstrap resampling method based on 100 replications. Percent error is calculated as follows: $\text{\textit{Percent error}}=\frac{\left|predicted-observed\right|}{observed}\times 100$}
\label{fig:Figure4}
\end{figure}

The small value of the errors indicates that all the parameters were very precisely estimated. The mixture proportions (the $\theta$'s) correspond exactly to the sample estimates shown in Table \ref{tab:Table1}, and the scale parameters (the $b$'s, $\beta$ and $\lambda$) reflect the changes over the period in both the median and the mean among the positive and negative values of real net wealth.\footnote{The correlation coefficients between the Weibull scale parameters ($\lambda$) and the two series of the median and mean net wealth values among the negatives are close to unity (0.982 and 0.955, respectively) and highly significant ($p\text{-value}<0.001$ in both cases). Similarly, the correlation coefficients between the values of the scale parameter of the Singh-Maddala ($b$), Dagum type I ($b$) and $\kappa$-generalized ($\beta$) distributions and the two series of the median and mean net wealth levels among the positives are all significant at the 1\% confidence level and equal, respectively, to 0.931, 0.983 and 0.998 for the median and 0.812, 0.925 and 0.935 for the mean.} The other parameters (the $a$'s, $\alpha$, $p$, $\kappa$, $q$ and $s$), characterizing distributional shape, are easiest to interpret by comparing predicted values for key distributional summary measures with their sample counterparts, as the effect of changing one of them is contingent on the value of the other parameters. For example, Figure \ref{fig:Figure4} shows that the overall mean net wealth and Gini coeffcient as estimated from the mixture models are very close to their sample estimates.\footnote{The analytic values for the mean and Gini coefficients, also reported in the last two columns of Table \ref{tab:Table2}, were obtained by substituting the estimated parameters into the relevant expressions given by Eqs. \eqref{eq:Equation13} and \eqref{eq:Equation14} with $r=1$ for the mean and Eqs. \eqref{eq:Equation25} for the Gini.}
%
%
However, the agreement (both in magnitude and temporal behavior) between the implied and sample estimates of the mean and Gini coefficient is much closer for the Singh-Maddala and $\kappa$-generalized mixture models than for the Dagum one. The mean and Gini coefficient associated with the latter model are in fact above the 95\% upper confidence limit of their corresponding sample estimates in six (from 1989 to 2005) and three (1994, 2003 and 2005) cases out of 9, respectively, and their percent error turns out to be relatively large compared to the other models\textemdash save for 1994, where both the mean and Gini predictions exhibited the lowest error, and 2009 with respect to the Gini coefficient implied by the Singh-Maddala mixture model. Overall, judging by the percent error of the mean and Gini coefficient of the net wealth distributions estimated from the three mixture models, the performance of the $\kappa$-generalized mixture model is appreciably superior to the other ones over most of the time span investigated.
%
\begin{figure}[!t]
\centering
\includegraphics[width=1.00\textwidth]{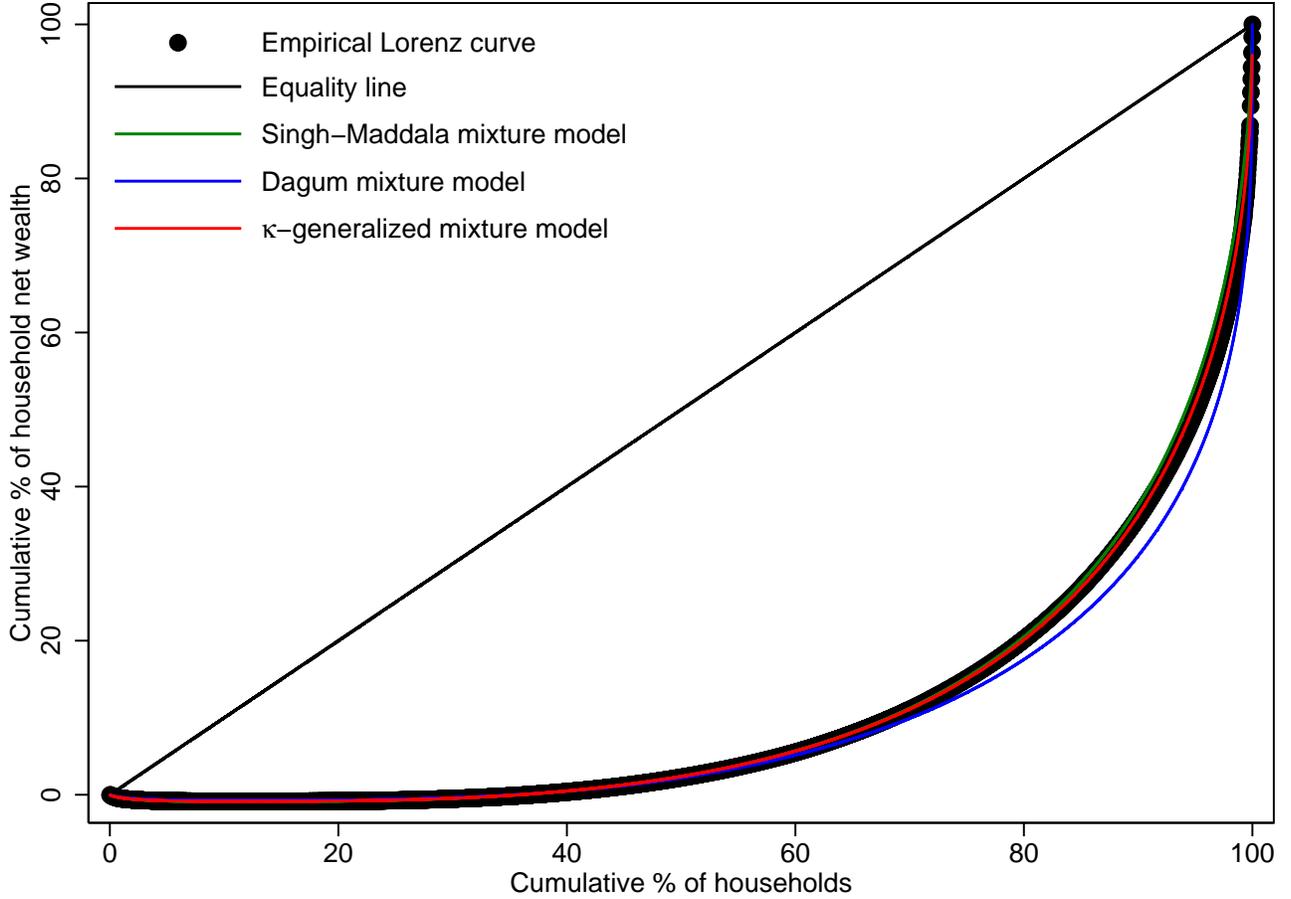}
\caption{Observed and calculated Lorenz curves for the U.S. household net wealth in 2003}
\label{fig:Figure5}
\end{figure}

The results summarized in the tables above allow us to emphasize a distinctive feature of wealth distributions, i.e. the concentration of density mass at zero. There is often a marked spike at zero because a relatively large fraction of the population has no wealth. Similar spikes do not occur with income distributions, where it is often the case that the density mass vanishes when income goes to zero. As the Weibull, Singh-Maddala, Dagum type I and $\kappa$-generalized distributions are zero-modal with a pole at the origin if, respectively, $s<1$, $a\leq1$, $ap\leq1$ and $\alpha\leq1$, it is easily verified from the estimates of these parameters that the probability density functions of the three mixture models inevitably transfer some density mass from the neighbouring values to the cited spike at zero\textemdash i.e. they diverge, rather than vanish, when the argument goes to zero from both the negative and positive ends of the wealth range.\footnote{The behavior around the mode of Weibull, Singh-Maddala and Dagum type I distributions is reviewed, e.g., in \cite{KleiberKotz2003}. For the $\kappa$-generalized distribution see \cite{ClementiDiMatteoGallegatiKaniadakis2008,ClementiGallegatiKaniadakis2009,ClementiGallegatiKaniadakis2010,ClementiGallegatiKaniadakis2012}.} This finding of a divergent probability density in the limit of zero wealth is also shared by other studies on the distribution of wealth (e.g. \cite{DragulescuYakovenko2001}).

The parameter estimates reported in Table \ref{tab:Table2} were also used to build estimated Lorenz curves by applying Eqs. \eqref{eq:Equation19}. The curves for 2003 are presented in Figure \ref{fig:Figure5} together with the empirical Lorenz curve estimate. Even if it is small, one can see a difference between the three predictions, in that the Lorenz curve estimated from the Dagum mixture model lies below the empirical one for approximately the top 30\% of the wealthiest households, while the Singh-Maddala and $\kappa$-generalized mixture models lead to estimated Lorenz curves exhibiting a degree of inequality that is much more in line with the observed one. In particular, the mean absolute difference between the empirical Lorenz data and the predicted values (averaged from all the survey years) amount to 0.004, 0.007 and 0.002, respectively, for the Dagum, Singh-Maddala and $\kappa$-generalized mixture models, thus indicating once again that the latter model gives a better match to the observed data than the other two.

It is interesting to note that the $\kappa$-generalized mixture model provides a better fit to most of the data than any of the alternative models regardless of the criterion used for comparison. For instance, by inspection of AIC and BIC values reported in the fourth- and third-last columns of Table \ref{tab:Table2}, it emerges that both the selection criteria agree on the $\kappa$-generalized mixture model as the preferred one for all of the survey waves.\footnote{Model selection criteria such as the Akaike \cite{Akaike1973} and Bayesian \cite{Schwarz1978} information criteria (AIC and BIC) will select, when comparing models with the same number of parameters, the model with the smallest log-likelihood value according to the formula $\left(2\times\text{\it logLik}\right)+\left(d\times npar\right)$, where $npar$ represents the number of parameters in the fitted model, and $d=2$ for the usual AIC or $d=\ln N$ ($N$ being the number of observations) for the so-called BIC.} To see if these differences in the performance of the alternative specifications are statistically significant, we adopt the Vuong approach to model selection \cite{Vuong1989}. This approach sets the model selection criterion in a hypothesis testing framework. More specifically, it tests the null hypothesis that the models under consideration are equidistant from a unknown ``true'' model against the alternative hypothesis that one model is closer. The test statistic is asymptotically normal under the null hypothesis and is quite straightforward to compute. Table \ref{tab:Table4} shows the results of the comparisons for the three mixture models.
%
\begin{table}[!t]
\centering
\begin{threeparttable}
\caption{Vuong test for model selection, 1984--2009\tnote{a}}
\newcolumntype{R}{>{\raggedleft\arraybackslash}X}%
\small
\begin{tabularx}{\textwidth}{lRR@{\hspace{0pt}}llRR@{\hspace{0pt}}l}
\hline\hline
\multirow{2}{*}{Wave}&\multicolumn{3}{c}{$\text{SM}$ vs. $\kappa\text{-gen}$}&&\multicolumn{3}{c}{$\text{D}$ vs. $\kappa\text{-gen}$}\\
\cline{2-4}\cline{6-8}
&Statistic&\multicolumn{2}{R}{$p\text{-value}$}&&Statistic&\multicolumn{2}{R}{$p\text{-value}$}\\
\hline
1984&-3.917&9e-05&\tnote{$\ast$}&&-0.159&0.874&\\
1989&-2.085&0.037&\tnote{$\ast$}&&-2.771&0.006&\tnote{$\ast$}\\
1994&-1.176&0.240&&&-2.966&0.003&\tnote{$\ast$}\\
1999&-1.741&0.082&&&-3.674&2e-04&\tnote{$\ast$}\\
2001&-2.665&0.008&\tnote{$\ast$}&&-2.481&0.013&\tnote{$\ast$}\\
2003&-2.486&0.013&\tnote{$\ast$}&&-2.529&0.011&\tnote{$\ast$}\\
2005&-2.121&0.034&\tnote{$\ast$}&&-2.336&0.019&\tnote{$\ast$}\\
2007&-2.434&0.015&\tnote{$\ast$}&&-2.462&0.014&\tnote{$\ast$}\\
2009&-1.964&0.050&\tnote{$\ast$}&&-2.675&0.007&\tnote{$\ast$}\\
\hline\hline
\end{tabularx}
\begin{tablenotes}
\footnotesize
\item\textit{Notes}: (a) SM = Singh-Maddala mixture model; D = Dagum mixture model; $\kappa$-gen = $\kappa$-generalized mixture model. The null hypothesis is that the competing models are equally close to the ``true'' data generating process. $^{\ast}$ Denotes 5\% statistical significance. 
\item\textit{Source}: Authors' own calculations using the PSID supplemental wealth files.
\end{tablenotes}
\label{tab:Table4}
\end{threeparttable}
\end{table}
%
As can be seen, if one takes the 5\% as the relevant significance level only in three cases (i.e. when comparing to the Singh-Maddala mixture model in the survey years 1994 and 1999 and to the Dagum one in 1984) the test concludes that the $\kappa$-generalized mixture model is observationally equivalent to its competitors, while in all the other cases (more than 83\% of all cases) its superiority as a descriptive model is found to be statistically significant.

The above evidence holds vis-\`a-vis a further check involving goodness of fit indicators such as the root mean squared error, defined as the square root of the average squared error between the observed and predicted values of the cumulative distribution function. In mathematical terms this is expressed as
\begin{equation}
RMSE=\sqrt{\frac{1}{N}\sum\limits^{N}_{i=1}\left[F_{\ast}\left(w_{i}\right)-\hat{F}_{N}\left(w_{i}\right)\right]^2},
\label{eq:Equation26}
\end{equation}
\sloppy where $F_{\ast}\left(w\right)$ is the distribution function deduced from the fitted mixture models and $\hat{F}_{N}\left(w\right)=\sum^{N}_{i=1}\pi_{i}\mathbf{1}_{A}\left(w\right)/\sum^{N}_{i=1}\pi_{i}$ denotes the empirical distribution function of the $N$ sample data ordered from lowest to highest carrying the $\pi_{i}$ along ($\mathbf{1}_{A}$ is the indicator function of the set $A=\left\{w|w_{i}\leq w\right\}$ and $\pi_{i}$ refers to the sampling weight of the $i$th observation). Clearly, lower values of $RMSE$ indicate a better fit. The comparison results between the competing models based on the above criterion are shown in Table \ref{tab:Table5}.
%
\begin{table}[!t]
\centering
\begin{threeparttable}
\caption{Goodness of fit comparisons for estimated mixture models of U.S. household net wealth, 1984--2009}
\newcolumntype{L}{>{\raggedright\arraybackslash}X}%
\newcolumntype{R}{>{\raggedleft\arraybackslash}X}%
\small
\begin{tabularx}{\textwidth}{LLrRRRR}
\hline\hline
Wave&Model&$RMSE$ ($\times 10^{2}$)&Rank&$A^{2}$ ($\times 10^{2}$)&$p\text{-value}$\tnote{a}&Rank\\
\hline
\multirow{3}{*}{1984}&SM&1.192&3&0.088&0.594&3\\
&D&0.986&1&0.054&0.743&1\\
&$\kappa$-gen&1.022&2&0.063&0.673&2\\
\hline
\multirow{3}{*}{1989}&SM&0.981&2&0.057&0.644&2\\
&D&1.118&3&0.064&0.594&3\\
&$\kappa$-gen&0.911&1&0.045&0.693&1\\
\hline
\multirow{3}{*}{1994}&SM&0.997&2&0.049&0.703&2\\
&D&1.080&3&0.056&0.693&3\\
&$\kappa$-gen&0.936&1&0.042&0.812&1\\
\hline
\multirow{3}{*}{1999}&SM&0.924&2&0.049&0.713&2\\
&D&1.058&3&0.055&0.614&3\\
&$\kappa$-gen&0.812&1&0.036&0.782&1\\
\hline
\multirow{3}{*}{2001}&SM&0.916&2&0.053&0.663&3\\
&D&1.008&3&0.050&0.733&2\\
&$\kappa$-gen&0.798&1&0.038&0.782&1\\
\hline
\multirow{3}{*}{2003}&SM&0.823&2&0.047&0.703&2\\
&D&0.947&3&0.050&0.673&3\\
&$\kappa$-gen&0.716&1&0.035&0.812&1\\
\hline
\multirow{3}{*}{2005}&SM&0.740&2&0.035&0.703&2\\
&D&0.882&3&0.041&0.653&3\\
&$\kappa$-gen&0.639&1&0.026&0.802&1\\
\hline
\multirow{3}{*}{2007}&SM&0.871&2&0.046&0.624&2\\
&D&0.923&3&0.047&0.713&3\\
&$\kappa$-gen&0.747&1&0.034&0.822&1\\
\hline
\multirow{3}{*}{2009}&SM&0.907&2&0.046&0.634&2\\
&D&0.992&3&0.048&0.663&3\\
&$\kappa$-gen&0.822&1&0.035&0.792&1\\
\hline\hline
\end{tabularx}
\begin{tablenotes}
\footnotesize
\item\textit{Notes}: (a) Upper tail $p$-value obtained by 100 bootstrap replications. The null hypothesis is that data come from the fitted Singh-Maddala (SM), Dagum (D) or $\kappa$-generalized ($\kappa$-gen) mixture model.
\item\textit{Source}: Authors' own calculations using the PSID supplemental wealth files.
\end{tablenotes}
\label{tab:Table5}
\end{threeparttable}
\end{table}
%
As can be seen, the $\kappa$-generalized mixture model of net wealth ranks first for all years but 1984, where it is outperformed by the Dagum mixture model.

Similar results are obtained by additionally performing an Anderson-Darling goodness of fit test that data come from the fitted Singh-Maddala, Dagum or $\kappa$-generalized mixture model. This test is known to be more powerful than other tests based on the empirical distribution function, since it provides equal sensitivity at the tails as at the median of the distribution \cite{Thode2002}.\footnote{The formula used for the test statistic is the one reported by \cite{Monahan2011}, which allows for weighted observations. Since the distribution of the Anderson-Darling test statistic is only known for data sets truly drawn from any given distribution \cite{Stephens1986}, while in our case the underlying distribution is itself determined by fitting to the data and hence varies from one data set to the next, the $p$-values for the test have been derived by making use of a nonparametric bootstrap method \cite{EfronTibshirani1993}. That is, given our $N$-vector of net wealth data, we generated 100 synthetic data sets by drawing new sequences of $N$ observations uniformly at random from the original data. We then fitted each synthetic data set individually to the three mixture models and calculated the test statistics for each one relative to its own models. Then we simply counted what fraction of the time each resulting statistic was larger than the value for the empirical data. This fraction is the $p$-value for each fit, and can be interpreted in the standard way: if it is larger than the chosen significance level, then the difference between the empirical data and the model can be attributed to statistical fluctuations alone; if it is smaller, the model is not a plausible fit to the data.} The last three columns of Table \ref{tab:Table5} report the test results for the nine sets of data. $p$-values are always larger than 0.05, meaning that (if one takes 5\% as the relevant significance level) in all cases the data can be statistically described by the three models. However, except for 1984, fitting the $\kappa$-generalized mixture model results both in lower values of the test statistic and higher $p$-values, thus offering superior performance over the Singh-Maddala and Dagum mixture models.

\begin{figure}[!t]
\centering
\includegraphics[width=1.00\textwidth]{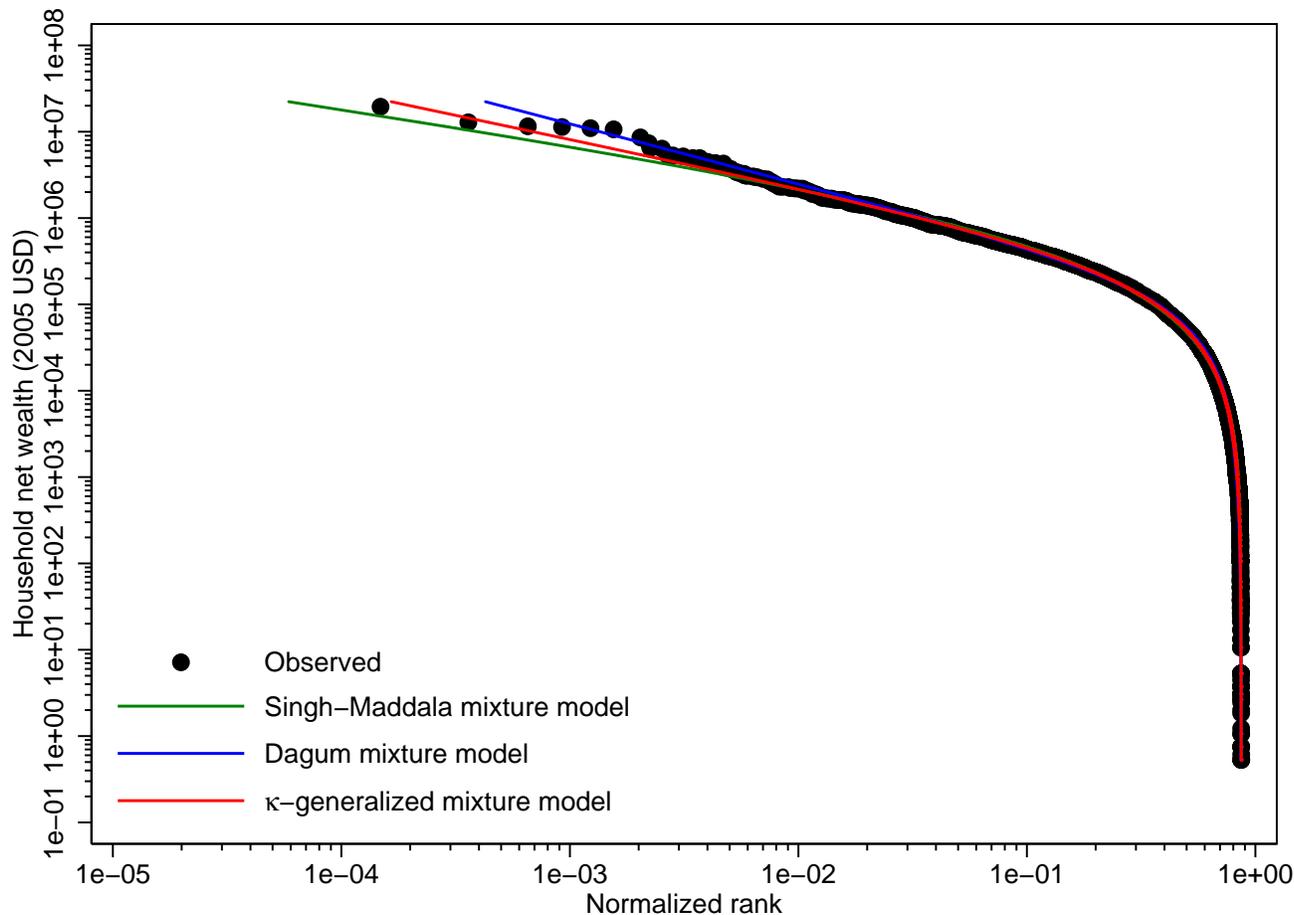}
\caption{Zipf plot for the positive values of U.S. household net wealth in 2003}
\label{fig:Figure6}
\end{figure}

Can these findings be ultimately ascribed to the different performance of the alternative densities used to characterize positive net wealth values? Figure \ref{fig:Figure6} presents for the 2003 PSID wave the relationship between log-rank and log-size along the positive support of the net wealth distribution.
%
%
This double-logarithmic framework, known as the Zipf plot, is natural to use when focusing on the top part of the distribution because it accentuates the upper tail, making it easier to detect deviations in that part of the distribution from the theoretical prediction of a particular model.\footnote{For an illustration of basic properties of the Zipf plot see e.g. \cite{StanleyBuldyrevHavlinMantegnaSalingerStanley1995}.} The lines show the predicted Zipf plots obtained from the fit of the models considered. As the figure reveals, all of them are in good agreement with the actual data in the low-middle range of the positive support of net wealth distribution. However, at the top tail there is a systematic departure of empirical observations from the theoretical predictions of the mixtures using the Singh-Maddala and Dagum type I specifications as descriptions of the positive net wealth values, while in the same part of the distributions the theoretical Zipf plot for the $\kappa$-generalized mixture model lies much closer to the empirical one.

This point is of particular relevance in the current context, both for the documented presence of long and fat tails towards the upper end of the U.S. net wealth distribution and the fact that all of the three densities accounting for the positive range of wealth obey the weak Pareto law \cite{Mandelbrot1960}. The weak version of the Pareto law states that the right-hand tail of a distribution behaves in the limit as a simple Pareto model, with an exponent that is a function of the parameters governing the shape of the distribution (see e.g. \cite{KramerZiebach2004} for an overview). The values of the Pareto index derived from parameter estimates of the Singh-Maddala, Dagum and $\kappa$-generalized mixture models are given in the sixth column of Table \ref{tab:Table2}.\footnote{See tables footnote (c) for formulas used to analytically derive the Pareto tail exponent in the three cases. For more details on the upper tail behavior of the $\kappa$-generalized distribution we refer the reader to \cite{ClementiGallegatiKaniadakis2009,ClementiGallegatiKaniadakis2010,ClementiGallegatiKaniadakis2012}. For the Singh-Maddala and Dagum distributions see instead \cite{WilflingKramer1993} and \cite{Kleiber1996,Kleiber2008}.} Remarkably, according to the $\kappa$-generalized mixture model the set of values for the index of the Pareto tail is closely in the narrow range $\left(1,2\right]$ that is generally found in empirical studies on the U.S. wealth distribution \cite[for instance]{LevySolomon1997,KlassBihamLevyMalcaiSolomon2006,KlassBihamLevyMalcaiSolomon2007}, whereas for the other two models the Paretian upper tail index oscillates systematically above (Singh-Maddala) and below (Dagum) the limits of this range.


\section{Summary and conclusions}
\label{sec:SummaryAndConclusions}

This paper mainly deals with the specification, analysis and application of models for net wealth distribution with support in the interval $\left(-\infty,\infty\right)$. These are mixtures\textemdash or, equivalently, convex representations\textemdash of three distributions with non-overlapping intervals, which have the advantage of providing a relatively flexible functional form and at the same time retain the advantages of parametric forms that are amenable to inference. The first distribution is a two-parameter Weibull model that describes the distribution of economic units with negative net wealth, i.e. covering the open interval $\left(-\infty,0\right)$; the second is a degenerate distribution with its unit mass concentrated at $w=0$; and the third is, alternatively, the three-parameter Singh-Maddala, Dagum type I or $\kappa$-generalized model that accounts for the distribution of economic units with positive net wealth, hence defined in the open interval $\left(0,\infty\right)$.

We have obtained closed formulas for the different probability functions, moments and standard tools for inequality measurement (i.e. the Lorenz curve and Gini concentration ratio). Except for the Dagum general model of net wealth \cite{Dagum1990,Dagum1994,Dagum2006a,Dagum2006b}, to the best of our knowledge this is the first time that the analytical properties of finite mixture models for net wealth based on alternative distributions to characterize positive values are fully derived.

The performance of the three mixture models has been checked against real data on U.S. household net wealth for different years. Goodness-of-fit comparisons reveal that all the three models are in good agreement with actual data, but the departure of empirical observations from the predictions of the Singh-Maddala and Dagum mixture models is always larger than in the case of the $\kappa$-generalized. In particular, the latter model suggests a superior fit in the right tail of data with respect to the others in many instances.

Finite mixture models deserve further attention in future. A feature of these models is that each of the parameters may be made a function of covariates summarizing household characteristics. Estimation of ``heterogeneous'' wealth distributions such as these, with distributional shape allowed to vary with personal characteristics, provides a route to decomposition analysis of the sources of differences in wealth inequality across years or countries.\footnote{For an application of parametric models in this manner, but to income, see \cite{BiewenJenkins2005,QuintanoDAgostino2006,BettiDAgostinoLemmi2008}.} This could be a good starting point for future research.


\bibliographystyle{elsarticle-num}
\bibliography{1209.4787v3}


\end{document}